\documentclass[12pt]{article}
\usepackage{amssymb}
\usepackage{graphicx}
\usepackage{amsmath}

\input{tcilatex}
\input{tcilatex}
\begin{document}

\title{Globally strongly convex cost functional for a coefficient inverse
problem }
\author{Larisa Beilina$^{\ast }$ and Michael V. Klibanov$^{\diamond }$ }
\date{}
\maketitle

\begin{abstract}
A Carleman Weight Function (CWF) is used to construct a new cost functional
for a Coefficient Inverse Problems for a hyperbolic PDE. Given a bounded set
of an arbitrary size in a certain Sobolev space, one can choose the
parameter of the CWF in such a way that the constructed cost functional will
be strongly convex on that set. Next, convergence of the gradient method,
which starts from an arbitrary point of that set, is established.\ Since
restrictions on the size of that set are not imposed, then this is the
global convergence.
\end{abstract}

\graphicspath{
{FIGURES/}
 {pics/}}

\footnote{$^{\ast }$Department of Mathematical Sciences, Chalmers University
of Technology and Gothenburg University, SE-42196 Gothenburg, Sweden, 
\texttt{\ larisa@chalmers.se}
\par
$^{\diamond }$ Department of Mathematics and Statistics University of North
Carolina at Charlotte, Charlotte, NC 28223, USA, \texttt{mklibanv{\@@}%
uncc.edu}}

\textbf{Key Words}: Coefficient inverse problem, wave-like equation,
Carleman weight function, strongly convex cost functional.

\textbf{AMS subject classifications.} 35R30, 35L05.

\section{Introduction}

\label{sec:1}

We consider a Coefficient Inverse Problem (CIP) for the equation $c\left(
x\right) u_{tt}=\Delta u,\left( x,t\right) \in \mathbb{R}^{3}\times \left(
0,T\right) $ with initial conditions $u\left( x,0\right) =f\left( x\right)
,u_{t}\left( x,0\right) =0$ \ and the unknown coefficient $c\left( x\right)
. $ Hence, the function $u=u\left( x,t,c\right) $ depends on the coefficient 
$c\left( x\right) $ nonlinearly. First, we derive an initial-boundary value
problem for a nonlinear integral differential equation with Volterra
integrals and with both Dirichlet and Neumann boundary conditions. The
coefficient $c\left( x\right) $ is not present in this equation. As soon as
the solution of this initial-boundary value problem is found, the function $%
c\left( x\right) $ can be easily calculated. To solve this problem, we
construct a weighted least squares Tikhonov-like functional for the latter
problem. The weight is the Carleman Weight Function (CWF), which is involved
in the Carleman estimate for the operator $c\left( x\right) \partial
_{t}^{2}-\Delta $. The main new result is Theorem 2 (subsection 2.3), which
ensures that, given a certain convex set of an arbitrary finite size in a
Hilbert space, one can choose the large parameter $\lambda >1$ of this CWF
in such a way that the above functional is strongly convex on this set.
Since restrictions on the size of that set are not imposed, we call this
\textquotedblleft global strong convexity".

To prove the strong convexity, we first prove a new Carleman estimate in
Theorem 1 (subsection 2.2). This estimate is derived for case when the
conventional term $\left( c\left( x\right) u_{tt}-\Delta u\right) ^{2}$ is
summed up with the nonlinear term $g\left( x,t\right) u\left( x,0\right)
u_{tt}\left( x,t\right) $ with a certain function $g\left( x,t\right) $. The
difficulty here is in the presence of the second derivative $u_{tt}\left(
x,t\right) $ in this term, since it is also present in the principal part of
the hyperbolic operator $c\left( x\right) \partial _{t}^{2}-\Delta .$ As far
as the authors are aware of, all currently known Carleman estimates are
obtained only for the case when the square of the linear principal part of a
PDE operator is involved.

The idea of obtaining a nonlinear integral differential equation without the
unknown coefficient present goes back to the method of proofs of global
uniqueness theorems for CIPs using Carleman estimates. This method was
originally proposed in \cite{BukhK}, also see, e.g. sections 1.10, 1.11 in 
\cite{BK} as well as surveys \cite{K,Y} and references cited there.\ 

We prove the global convergence of the gradient method of the minimization
of our functional.\ More precisely, we prove that this method converges to
the unique minimizer on the above set if starting from an arbitrary point of
that set. In addition, the distance between the minimizer and the exact
solution of the original problem is estimated in the case when the data are
given with an error. Also, keeping in mind future numerical studies, we
prove similar results for some finite dimensional approximations of that
integral differential equation. Finally, we outline an algorithm of solving
that minimization problem using the FEM. The latter might be useful in
computations. Numerical testing of these ideas would require a substantial
additional effort, which is outside of the scope of the current paper. We
plan to do this in the future.

The assumption that we work on a priori given bounded set is going along
well with the Tikhonov concept for ill-posed problems, see, e.g. section 1.4
of \cite{BK}. By this concept, one should seek for the solution of an
ill-posed on an a priori given bounded set, which can also be considered as
the set of admissible parameters.

A cost functional with the CWF in it, which is strongly convex on a bounded
convex set of an arbitrary size, was constructed for a similar CIP in the
earlier work of the second author \cite{Klib97}. However, the strong
convexity was established only in the case when a series resulting from the
separation of $x$ and $t$ variables was truncated. It is unclear how the
result of \cite{Klib97} would change if the number of terms in that
truncated series would tend to infinity. As a result of the separation of
variables, the CWF for the Laplace operator was used in \cite{Klib97}.
Compared with \cite{Klib97}, the main new element of this paper is that we
do not truncate any series in Theorem 2. In addition, we do not consider the
separation of $x$ and $t$ variables and use the CWF for the above hyperbolic
operator.

In the recent interesting work \cite{Bad} a different CWF was used to
construct a globally convergent numerical method for an analog of our CIP
for the case of the equation $v_{tt}=\Delta v-q(x)v$ with the unknown
coefficient $q(x).$ The method of \cite{Bad} requires the minimization on
each iterative step of a certain quadratic functional with the CWF in it. As
a result, a good approximation for $q(x)$ is obtained, if starting from any
point of a bounded set of an arbitrary radius (in a certain space). This
approach is different from ours in the sense that a nonlinear integral
differential equation was not obtained and a globally strongly convex cost
functional was not constructed in \cite{Bad}.

Conventional numerical methods for CIPs, such as, e.g. gradient method and
Newton method, are based on the minimization of conventional least squares
cost functionals for CIPs. These functionals suffer from the phenomenon of
local minima and ravines. Thus, these methods converge locally, i.e. their
convergence is guaranteed only if the starting point is located in a small
neighborhood of the solution. Recently a globally convergent method for our
CIP was developed analytically and verified computationally in a series of
publications of the authors, which were summarized in their book \cite{BK}.
This method is not using an optimization procedure and is significantly
different from the one of the current paper.

In some theorems below we analyze the case when an error in the data is
present. We assume that its level is sufficiently small. The smallness
assumption about the error is a natural one for an ill-posed problem.
Indeed, the intuition says that a numerical method cannot work well if the
data contain a large error. Another argument here is that it is well known
that the theory is usually more pessimistic than computations.\ For example,
the global convergence theorem 2.9.4 of \cite{BK} also works with a small
error in the data.\ Nevertheless, that method works well with experimental
data, which are very noisy: see Chapter 5 and section 6.9 of \cite{BK} as
well as \cite{BTKF} and references cited in these publications.

In section 2 we formulate our inverse problem and theorems. In sections 3-5
we prove these theorems. In section 6 we outline an algorithm which works
with finite elements.

\section{The Coefficient Inverse Problem and Main Results}

\label{sec:2}

Let $\Omega \subset \mathbb{R}^{3}$ be a convex bounded domain with a
piecewise-smooth boundary $\partial \Omega $.\ For any $T>0$ denote $%
Q_{T}=\Omega \times (0,T),S_{T}=\partial \Omega \times \left( 0,T\right) .$
Let the function $c\left( x\right) $ satisfies the following conditions%
\begin{equation}
1\leq c\left( x\right) \leq 1+b,\forall x\in \mathbb{R}^{3},  \label{2.1}
\end{equation}%
\begin{equation}
c\left( x\right) =\overline{b}=const.\in \left[ 1,1+b\right] ,x\in \mathbb{R}%
^{3}\diagdown \Omega ,  \label{2.2}
\end{equation}%
\begin{equation}
c\in C^{1}\left( \mathbb{R}^{3}\right) ,  \label{2.3}
\end{equation}%
where numbers $b,\overline{b}>0$ are given. We use \textquotedblleft 1" here
in (\ref{2.2}) for the normalization only. In addition, we assume that there
exists a point $x_{0}\in \mathbb{R}^{3}\diagdown \overline{\Omega }$ such
that%
\begin{equation}
\left( \nabla c,x-x_{0}\right) \geq 0,\forall x\in \overline{\Omega },
\label{2.4}
\end{equation}%
where $\left( ,\right) $ denotes the scalar product in $\mathbb{R}^{3}.$
Condition (\ref{2.4}) is imposed to guarantee the validity of the Carleman
estimate for the operator $c\left( x\right) \partial _{t}^{2}-\Delta ,$ see
Theorem 1.10.2 and Corollary 1.10.2 in \cite{BK}. Let $\Omega ^{\prime
}\subset \mathbb{R}^{3}$ be another bounded domain such that $\Omega \subset
\Omega ^{\prime },\partial \Omega \cap \partial \Omega ^{\prime
}=\varnothing .$ We assume that the function $f\left( x\right) $ satisfies
the following conditions 
\begin{equation}
f\in H^{7}\left( \mathbb{R}^{3}\right) ,  \label{2.6}
\end{equation}%
\begin{equation}
f\left( x\right) =0\text{ for }x\in \mathbb{R}^{3}\diagdown \Omega ^{\prime
}.  \label{2.61}
\end{equation}%
The embedding theorem, (\ref{2.6}) and (\ref{2.61}) imply that $f\in
C^{5}\left( \mathbb{R}^{3}\right) .$ In addition, we assume below that 
\begin{equation}
\Delta f\geq \xi =const.>0\text{ for }x\in \overline{\Omega }\text{.}
\label{2.104}
\end{equation}%
A discussion of condition (\ref{2.104}) can be found in subsection 2.4.

\subsection{Coefficient Inverse Problem}

\label{sec:2.1}

Consider the following Cauchy problem%
\begin{equation}
c\left( x\right) u_{tt}=\Delta u,\left( x,t\right) \in \mathbb{R}^{3}\times
\left( 0,T\right) ,  \label{2.7}
\end{equation}%
\begin{equation}
u\left( x,0\right) =f\left( x\right) ,u_{t}\left( x,0\right) =0.  \label{2.8}
\end{equation}%
It follows from corollary 4.2 of Chapter 4 of the book \cite{Lad} that
conditions (\ref{2.1})-(\ref{2.3}), (\ref{2.6}) and (\ref{2.61}) guarantee
uniqueness and existence of the solution $u\in H^{2}\left( \mathbb{R}%
^{3}\times \left( 0,T\right) \right) $ of the problem (\ref{2.7}), (\ref{2.8}%
). To apply our technique, we need $u_{tt}\in C^{3}\left( \mathbb{R}%
^{3}\times \left[ 0,T\right] \right) .$ Since by the embedding theorem $%
H^{6}\left( \mathbb{R}^{3}\times \left( 0,T\right) \right) \subset
C^{3}\left( \mathbb{R}^{3}\times \left[ 0,T\right] \right) ,$ then we assume
that 
\begin{equation}
u\in H^{8}\left( \mathbb{R}^{3}\times \left( 0,T\right) \right) .
\label{2.9}
\end{equation}%
Hence, the trace theorem justifies the smoothness (\ref{2.6}) for $f\left(
x\right) =u\left( x,0\right) .$ Using results of Chapter 4 of \cite{Lad},
one can show that some smoothness conditions being imposed on functions $c,f$
guarantee (\ref{2.9}). However, we leave aside these conditions for brevity
and just assume (\ref{2.9}) below. Note that the issue of the minimal
smoothness of coefficients is rarely a concern in the theory of CIPs, since
these problems are quite difficult ones even for sufficiently smooth
coefficients, see, e.g. \cite{Nov1,Nov2}.

Equation (\ref{2.7}) is the acoustics equation, where $c^{-1/2}\left(
x\right) $ is the speed of sound and $u\left( x,t\right) $ is the pressure
of the acoustic wave at the point $x$ at the moment of time $t$. In the 2-d
case (\ref{2.7}) can be derived from the Maxwell's equations.\ In this case $%
c\left( x\right) $ is the spatially distributed dielectric constant and $%
u\left( x,t\right) $ is the amplitude of one of components of the electric
wave. In addition, (\ref{2.7}) was successfully used in \cite{BK,BTKF} to
model the propagation in $\mathbb{R}^{3}$ of one of components of the
electric field in the case of experimental data with the same interpretation
of functions $c\left( x\right) $ and $u\left( x,t\right) $ as above$.$ A
numerical explanation of the latter can be found in \cite{BM}.

\textbf{Coefficient Inverse Problem (CIP). }Suppose that conditions (\ref%
{2.1})-(\ref{2.9}) are satisfied. Assume that the coefficient $c(x)$ is
unknown inside of the domain $\Omega $. Determine the function $c(x)$ for $%
x\in \Omega ,$ assuming that the following function $s(x,t)$ is known 
\begin{equation}
u|_{S_{T}}=s(x,t).  \label{2.11}
\end{equation}

The function $s(x,t)$ in (\ref{2.11}) models the boundary measurement.
Having $s(x,t)$, one can uniquely solve the initial boundary value problem
for equation (\ref{2.7}) with the initial condition (\ref{2.8}) outside of
the domain $\Omega .$ Hence, the normal derivative is known, 
\begin{equation}
\partial _{\nu }u\mid _{S_{T}}=p\left( x,t\right) .  \label{2.13}
\end{equation}

\textbf{Remark 1}. It would be sufficient to know the function $p\left(
x,t\right) $ in (\ref{2.13}) only on a part of the boundary $\partial \Omega
.$ To extend our method to this case, one should specify a part of the
Carleman estimate of Theorem 1, see, e.g. \cite{Bad} for such a
specification.\ We do not follow this root for brevity. Uniqueness of this
CIP \ under conditions (\ref{2.4}), (\ref{2.104}) was established via the
method of \cite{BukhK}, see, e.g. theorem 1.10.5.1 in \cite{BK} and theorem
3.1 in \cite{K}.

\subsection{The Carleman estimate with a nonlinear term}

\label{sec:2.2}

Choose a point $x_{0}\in \mathbb{R}^{3}\diagdown \overline{\Omega }.$ Let
the number $\eta \in \left( 0,1\right) $ and let $\lambda >1$ be a large
parameter. Consider functions $\psi ,\varphi _{\lambda }$,%
\begin{equation*}
\psi \left( x,t\right) =\left\vert x-x_{0}\right\vert ^{2}-\eta
t^{2},\varphi _{\lambda }\left( x,t\right) =\exp \left( \lambda \psi \left(
x,t\right) \right) .
\end{equation*}%
For any number $\eta \in \left( 0,1\right) $ one can choose a sufficiently
large $T$ such that 
\begin{equation}
N=N\left( \Omega ,x_{0},\eta ,T\right) =\eta T^{2}-\max_{x\in \overline{%
\Omega }}\left\vert x-x_{0}\right\vert ^{2}>0.  \label{101}
\end{equation}%
Choose a number $d\in \left( 0,\min_{x\in \overline{\Omega }}\left\vert
x-x_{0}\right\vert ^{2}\right) .$ Hence, $\Omega \subset \left\{ \left\vert
x-x_{0}\right\vert ^{2}>d\right\} $. Consider the domain $P_{d}$ and the
number $M>0,$%
\begin{equation*}
P_{d}=\left\{ \left( x,t\right) \in Q_{T}:\psi \left( x,t\right) >d\right\}
,M=\max_{\overline{Q}_{T}}\psi \left( x,t\right) =\max_{x\in \overline{%
\Omega }}\left\vert x-x_{0}\right\vert ^{2}.
\end{equation*}%
By (\ref{101}) $\overline{P}_{d}\cap \left\{ t=T\right\} =\varnothing .$ In
fact, $P_{d}$ is a part of the hyperboloid $\left\{ \left( x,t\right)
:\left\vert x-x_{0}\right\vert ^{2}-\eta t^{2}>d\right\} .$ Denote $\Omega
_{T}=\left\{ \left( x,t\right) :x\in \Omega ,t=T\right\} .$ Let the function 
$g\left( x,t\right) $ be such that $g,\partial _{t}g\in C\left( \overline{Q}%
_{T}\right) .$

\textbf{Theorem 1.} \emph{Assume that the function }$c\left( x\right) $\emph{%
\ satisfies conditions (\ref{2.1}), (\ref{2.3}) and (\ref{2.4}).\ Then there
exist a sufficiently small number }$\eta =\eta \left( \Omega
,x_{0},\left\Vert c\right\Vert _{C^{1}\left( \overline{\Omega }\right)
}\right) \in \left( 0,1\right) $\emph{, a positive number }$%
C_{1}=C_{1}\left( \eta ,b,\left\Vert g\right\Vert _{C\left( \overline{Q}%
_{T}\right) },\left\Vert \partial _{t}g\right\Vert _{C\left( \overline{Q}%
_{T}\right) }\right) $\emph{\ and a sufficiently large number }

$\overline{\lambda }=\overline{\lambda }\left( \eta ,T,b,\left\Vert
g\right\Vert _{C\left( \overline{Q}_{T}\right) },\left\Vert \partial
_{t}g\right\Vert _{C\left( \overline{Q}_{T}\right) }\right) >1$\emph{, all
three numbers depending only on listed parameters, such that if }$T=T\left(
\eta ,x_{0},\Omega \right) >0$ \emph{is so large that (\ref{101}) holds,
then for all }$\lambda \geq \overline{\lambda }$\emph{\ the following
Carleman estimate is valid}%
\begin{equation*}
\int\limits_{Q_{T}}\left( c\left( x\right) u_{tt}-\Delta u\right)
^{2}\varphi _{\lambda }^{2}dxdt+\int\limits_{Q_{T}}g\left( x,t\right)
u\left( x,0\right) u_{tt}\left( x,t\right) \varphi _{\lambda }^{2}dxdt
\end{equation*}%
\begin{equation*}
+C_{1}\lambda ^{3}\exp \left( 2\lambda M\right) \left( \left\Vert \partial
_{\nu }u\mid _{S_{T}}\right\Vert _{L_{2}\left( S_{T}\right) }^{2}+\left\Vert
u\right\Vert _{H^{1}\left( S_{T}\right) }^{2}\right)
\end{equation*}%
\begin{equation}
+C_{1}\lambda ^{3}\exp \left( -2\lambda N\right) \left( \left\Vert
u_{t}\right\Vert _{L_{2}\left( \Omega _{T}\right) }^{2}+\left\Vert
u\right\Vert _{H^{1}\left( \Omega _{T}\right) }^{2}\right) +C_{1}\exp \left(
-\lambda N\right) \left\Vert u_{t}\right\Vert _{L_{2}\left( Q_{T}\right)
}^{2}  \label{2.131}
\end{equation}%
\begin{equation*}
\geq C_{1}\int\limits_{Q_{T}}\left( \lambda u_{t}^{2}+\lambda \left( \nabla
u\right) ^{2}+\lambda ^{3}u^{2}\right) \varphi _{\lambda }^{2}dxdt,
\end{equation*}%
\begin{equation*}
\forall u\in \left\{ U\in H^{2}\left( Q_{T}\right) :U_{t}\left( x,0\right)
=0\right\} .
\end{equation*}%
\emph{In the case }$c\left( x\right) \equiv 1$\emph{\ one can choose any
number }$\eta \in \left( 0,1\right) .$

\textbf{Remarks 2:}

\textbf{1}. The main difficulty of the proof of this theorem (section 3) is
due to the presence of the second derivative $u_{tt}$ in the nonlinear term $%
g\left( x,t\right) u_{tt}\left( x,t\right) u\left( x,0\right) $, since this
derivative is also a part of the operator $c\left( x\right) \partial
_{t}^{2}-\Delta $. On the other hand, $\left( c\left( x\right) u_{tt}-\Delta
u\right) ^{2}$ in (\ref{2.131}) is the standard term in the Carleman
estimate for this operator, see, e.g. Theorem 1.10.2 and Corollary 1.10.2 of 
\cite{BK}. The only non-standard element here is the absence of the integral
over $\overline{Q}_{T}\cap \left\{ t=0\right\} .$ This absence is due to the
fact that formulae (1.86) and (1.87) in \cite{BK} imply that the
corresponding integral over $\overline{Q}_{T}\cap \left\{ t=0\right\} $
equals zero, since $u_{t}\left( x,0\right) =0$.

\textbf{2}. If the nonlinear term $g\left( x,t\right) u\left( x,0\right)
u_{tt}\left( x,t\right) $ would be absent, then we would not need terms in
the third line of (\ref{2.131}), and the fourth line would be replaced
simply with $\left\Vert u\right\Vert _{H^{1}\left( Q_{T}\right) }^{2}.$ This
follows from a slight modification of the technique, which was proposed for
the first time in \cite{KlibM}. This technique can also be found in, e.g.
theorem 5.1 of the more recent publication \cite{K}, also, see references in
section 5.5 of \cite{K}.

\subsection{Strong convexity}

\label{sec:2.3}

In this subsection we formulate our main result. Denote 
\begin{equation}
\overline{s}\left( x,t\right) =\partial _{t}^{2}s\left( x,t\right) ,%
\overline{p}\left( x,t\right) =\partial _{t}^{2}p\left( x,t\right) .
\label{2.140}
\end{equation}%
Let $\widetilde{w}=u_{tt}.$ Then by (\ref{2.9}) $\widetilde{w}\in
H_{0}^{6}\left( Q_{T}\right) .$ Hence, (\ref{2.7}), (\ref{2.8}) and (\ref%
{2.140}) imply that 
\begin{equation}
c\left( x\right) =\frac{\left( \Delta f\right) \left( x\right) }{\widetilde{w%
}\left( x,0\right) },x\in \Omega ,  \label{2.15}
\end{equation}
\begin{equation}
\frac{\left( \Delta f\right) \left( x\right) }{\widetilde{w}\left(
x,0\right) }\widetilde{w}_{tt}-\Delta \widetilde{w}=0,\left( x,t\right) \in
Q_{T},  \label{2.16}
\end{equation}%
\begin{equation}
\widetilde{w}_{t}\left( x,0\right) =0,\widetilde{w}\mid _{S_{T}}=\overline{s}%
\left( x,t\right) ,\partial _{\nu }\widetilde{w}\mid _{S_{T}}=\overline{p}%
\left( x,t\right) .  \label{2.17}
\end{equation}%
We now want to obtain zero boundary conditions at $S_{T}.$ To do this,
assume that there exists a function $F\left( x,t\right) $ such that%
\begin{equation}
F\in H^{6}\left( Q_{T}\right) ,F_{t}\left( x,0\right) =0,F\mid _{S_{T}}=%
\overline{s}\left( x,t\right) ,\partial _{\nu }F\mid _{S_{T}}=\overline{p}%
\left( x,t\right) .  \label{2.18}
\end{equation}%
Denote $w=\widetilde{w}-F.$ Hence, by (\ref{2.15})%
\begin{equation}
c\left( x\right) =\frac{\left( \Delta f\right) \left( x\right) }{\left(
w+F\right) \left( x,0\right) }.  \label{2.151}
\end{equation}%
For any integer $s\geq 2$ denote%
\begin{equation}
H_{0}^{s}\left( Q_{T}\right) =\left\{ u\in H^{s}\left( Q_{T}\right)
:u_{t}\left( x,0\right) =0,u\mid _{S_{T}}=0,\partial _{\nu }u\mid
_{S_{T}}=0\right\} .  \label{2.181}
\end{equation}%
Then (\ref{2.16})-(\ref{2.181}) imply that%
\begin{equation}
\frac{\left( \Delta f\right) \left( x\right) }{\left[ w\left( x,0\right)
+F\left( x,0\right) \right] }\left( w+F\right) _{tt}-\Delta \left(
w+F\right) =0,\left( x,t\right) \in Q_{T},  \label{2.182}
\end{equation}%
\begin{equation}
w_{t}\left( x,0\right) =0,w\mid _{S_{T}}=0,\partial _{\nu }w\mid _{S_{T}}=0,%
\text{ i.e. }w\in H_{0}^{6}\left( Q_{T}\right) .  \label{2.183}
\end{equation}%
\ We focus below on the solution of the problem (\ref{2.182}), (\ref{2.183}%
). Indeed, if an approximate solution of this problem is found, then the
corresponding approximation for the target coefficient $c\left( x\right) $
can be found via (\ref{2.15}) where $\widetilde{w}=w+F$. Note that equation (%
\ref{2.182}) is nonlinear with respect to the function $w$.

We need the smoothness $w\in H_{0}^{6}\left( Q_{T}\right) $ because the
proof of Theorem 2 uses the fact that three derivatives of the function $w$
are bounded.\ This can be ensured if, e.g. $w\in C^{3}\left( \overline{Q}%
_{T}\right) .$ Indeed, by the embedding theorem%
\begin{equation}
H^{6}\left( Q_{T}\right) \subset C^{3}\left( \overline{Q}_{T}\right)
,\left\Vert y\right\Vert _{C^{3}\left( \overline{Q}_{T}\right) }\leq
C_{2}\left\Vert y\right\Vert _{H^{6}\left( Q_{T}\right) },\forall y\in
H^{6}\left( Q_{T}\right) ,  \label{2.19}
\end{equation}%
where the number $C_{2}=C_{2}\left( Q_{T}\right) >0$ depends only on the
domain $Q_{T}.$ Also, by the trace theorem 
\begin{equation}
\left\Vert y\left( x,0\right) \right\Vert _{H^{5}\left( \Omega \right) }\leq
C_{2}\left\Vert y\right\Vert _{H^{6}\left( Q_{T}\right) },\forall y\in
H^{6}\left( Q_{T}\right) .  \label{2.190}
\end{equation}%
Let $R>0$ be an arbitrary number such that 
\begin{equation}
\left\Vert \Delta f\right\Vert _{H^{5}\left( \Omega \right) }\leq C_{2}R.
\label{2.20}
\end{equation}%
For every function $v\in H_{0}^{6}\left( Q_{T}\right) $ such that $\left(
v+F\right) \left( x,0\right) >0$ in $\overline{\Omega }$ denote%
\begin{equation}
A\left( v\right) =\frac{\left( \Delta f\right) \left( x\right) }{v\left(
x,0\right) +F\left( x,0\right) },x\in \Omega .  \label{2.21}
\end{equation}

Consider the set of functions $G\left( Q_{T},b,R,f,F\right) $ defined as%
\begin{equation}
G=G\left( Q_{T},b,R,f,F\right) =\left\{ 
\begin{array}{c}
v\in H_{0}^{6}\left( Q_{T}\right) , \\ 
\left\Vert v\right\Vert _{H^{6}\left( Q_{T}\right) }\leq R, \\ 
\left( 1+b\right) ^{-1}\left( \Delta f\right) \left( x\right) \leq v\left(
x,0\right) +F\left( x,0\right) \leq \left( \Delta f\right) \left( x\right) 
\text{ in }\overline{\Omega }, \\ 
\left( \nabla A\left( v\right) \left( x\right) ,x-x_{0}\right) \geq 0\text{
in }\overline{\Omega }.%
\end{array}%
\right.  \label{2.22}
\end{equation}%
Denote $Int\left( G\right) $ the open set of interior points of $G$.
Inequality (\ref{2.20}) guarantees that the function $\left( \Delta f\right)
\left( x\right) $ is in the proper range. Indeed, $\left\Vert v\left(
x,0\right) +F\left( x,0\right) \right\Vert _{H^{5}\left( \Omega \right)
}\leq C_{2}R$ in this case: by (\ref{2.190}) and (\ref{2.22}). If $c=c\left(
x,v\right) =A\left( v\right) ,v\in G,$ then by (\ref{2.104}), (\ref{2.190})
and (\ref{2.20}) the function $c$ satisfies conditions (\ref{2.1}), (\ref%
{2.3}), (\ref{2.4}).

We need Proposition 1, since convex functionals are typically defined on
convex sets.

\textbf{Proposition 1}. $G$ \emph{is a convex set.}

\textbf{Proof}. Let $\beta \in \left[ 0,1\right] $ be an arbitrary number
and let $v_{1},v_{2}\in G$ be two arbitrary functions. We should prove that
the function $\overline{v}=\beta v_{1}+\left( 1-\beta \right) v_{2}\in G.$
Obviously this function satisfies conditions of first three lines of (\ref%
{2.22}). Consider now the fourth line. By (\ref{2.21}) 
\begin{equation*}
\left( \nabla A\left( \overline{v}\right) \left( x\right) ,x-x_{0}\right) =%
\frac{\beta }{\left( \overline{v}+F\right) ^{2}\left( x,0\right) }\left(
\left( \nabla \left( \Delta f\right) \left( v_{1}+F\right) \left( x,0\right)
-\Delta f\nabla \left( v_{1}+F\right) \left( x,0\right) \right)
,x-x_{0}\right)
\end{equation*}%
\begin{equation*}
+\frac{\left( 1-\beta \right) }{\left( \overline{v}+F\right) ^{2}\left(
x,0\right) }\left( \left( \nabla \left( \Delta f\right) \left(
v_{2}+F\right) \left( x,0\right) -\Delta f\nabla \left( v_{2}+F\right)
\left( x,0\right) \right) ,x-x_{0}\right) .
\end{equation*}%
Since $\left( \nabla A\left( v_{j}\right) ,x-x_{0}\right) \geq 0$ for $%
j=1,2, $ then 
\begin{equation*}
\left( \left( \nabla \left( \Delta f\right) \left( v_{j}+F\right) \left(
x,0\right) -\Delta f\nabla \left( v_{j}+F\right) \left( x,0\right) \right)
,x-x_{0}\right) \geq 0.
\end{equation*}
Hence, $\left( \nabla A\left( \overline{v}\right) \left( x\right)
,x-x_{0}\right) \geq 0.$ $\square $

We now reformulate the problem (\ref{2.182}), (\ref{2.183}) as: \emph{Find a
function }$w\left( x,t\right) $\emph{\ such that}%
\begin{equation}
Y\left( w\right) :=A\left( w\right) \left( w+F\right) _{tt}-\Delta \left(
w+F\right) =0,\left( x,t\right) \in Q_{T},w\in G.  \label{2.184}
\end{equation}%
Let $\alpha \in \left( 0,1\right) $ be the regularization parameter. To
solve the problem (\ref{2.184}), we construct the following weighted
Tikhonov regularization functional $J_{\lambda ,\alpha }\left( w\right)
:G\rightarrow \mathbb{R}$, 
\begin{equation}
J_{\lambda ,\alpha }\left( w\right) =\int\limits_{Q_{T}}\left[ A\left(
w\right) \left( w+F\right) _{tt}-\Delta \left( w+F\right) \right]
^{2}\varphi _{\lambda }^{2}dxdt+\alpha \left\Vert w\right\Vert _{H^{6}\left(
Q_{T}\right) }^{2},w\in G,  \label{2.23}
\end{equation}

\textbf{Theorem 2}. \emph{Let }$\eta $\emph{\ and }$T$\emph{\ be numbers of
Theorem 1 and let }$Int\left( G\right) \neq \varnothing $\emph{. Let }$%
J_{\lambda ,\alpha }^{\prime }\left( w\right) \left( h\right) ,\forall h\in
H_{0}^{6}\left( Q_{T}\right) $\emph{\ be the Fr\'{e}chet derivative of the
functional }$J_{\lambda ,\alpha }$\emph{\ at the point }$w\in Int\left(
G\right) .$ \emph{Then there exists a constant }$C=C\left( \eta
,T,b,R,G\right) >0$\emph{\ and a sufficiently large number }$\lambda
_{0}=\lambda _{0}\left( \eta ,T,b,R,G\right) >1,$ \emph{both depending only
on listed parameters, such that for all }$\lambda \geq \lambda _{0}$\emph{\
and for all }$\alpha \geq 2C\exp \left( -\lambda N\right) $ \emph{the
functional }$J_{\lambda ,\alpha }\left( w\right) $\emph{\ is strongly convex
on the set }$G.$\emph{\ More precisely} \emph{\ }%
\begin{equation*}
J_{\lambda ,\alpha }\left( w_{2}\right) -J_{\lambda ,\alpha }\left(
w_{1}\right) -J_{\lambda ,\alpha }^{\prime }\left( w_{1}\right) \left(
w_{2}-w_{1}\right)
\end{equation*}%
\begin{equation}
\geq C\int\limits_{Q_{T}}\left[ \lambda \left( \partial _{t}\left(
w_{2}-w_{1}\right) \right) ^{2}+\lambda \left( \nabla \left(
w_{2}-w_{1}\right) \right) ^{2}+\lambda ^{3}\left( w_{2}-w_{1}\right) ^{2}%
\right] \varphi _{\lambda }^{2}dxdt  \label{2.24}
\end{equation}%
\begin{equation*}
+\frac{\alpha }{2}\left\Vert \left( w_{2}-w_{1}\right) \right\Vert
_{H^{6}\left( Q_{T}\right) }^{2}\geq \frac{\alpha }{2}\left\Vert
w_{2}-w_{1}\right\Vert _{H^{6}\left( Q_{T}\right) }^{2},\forall w_{1}\in
Int\left( G\right) ,\forall w_{2}\in G.
\end{equation*}%
\emph{In particular, (\ref{2.24}) implies that} 
\begin{equation*}
J_{\lambda ,\alpha }\left( w_{2}\right) -J_{\lambda ,\alpha }\left(
w_{1}\right) -J_{\lambda ,\alpha }^{\prime }\left( w_{1}\right) \left(
w_{2}-w_{1}\right)
\end{equation*}%
\begin{equation}
\geq C\exp \left( 2\lambda d\right) \left\Vert w_{2}-w_{1}\right\Vert
_{H^{1}\left( P_{d}\right) }^{2}+\frac{\alpha }{2}\left\Vert
w_{2}-w_{1}\right\Vert _{H^{6}\left( Q_{T}\right) }^{2},\forall w_{1}\in
Int\left( G\right) ,\forall w_{2}\in G.  \label{2.242}
\end{equation}

\textbf{Corollary 1}. \emph{Let }%
\begin{equation}
\widetilde{J}_{\lambda }\left( w\right) =\int\limits_{Q_{T}}\left[ A\left(
w\right) \left( w+F\right) _{tt}-\Delta \left( w+F\right) \right]
^{2}\varphi _{\lambda }^{2}xdxdt.  \label{2.241}
\end{equation}%
\emph{\ Then} \emph{\ } 
\begin{equation*}
\widetilde{J}_{\lambda }\left( w_{2}\right) -\widetilde{J}_{\lambda }\left(
w_{1}\right) -\widetilde{J}_{\lambda }^{\prime }\left( w_{1}\right) \left(
w_{2}-w_{1}\right) \geq C\exp \left( 2\lambda d\right) \left\Vert
w_{2}-w_{1}\right\Vert _{H^{1}\left( P_{d}\right) }^{2}
\end{equation*}%
\begin{equation}
-C\lambda ^{3}\exp \left( -2\lambda N\right) \left( \left\Vert \partial
_{t}\left( w_{2}-w_{1}\right) \right\Vert _{L_{2}\left( \Omega _{T}\right)
}^{2}+\left\Vert w_{2}-w_{1}\right\Vert _{H^{1}\left( \Omega _{T}\right)
}^{2}\right)  \label{2.243}
\end{equation}%
\begin{equation*}
-C\exp \left( -\lambda N\right) \left\Vert w_{2}-w_{1}\right\Vert
_{L_{2}\left( Q_{T}\right) }^{2},\forall w_{1}\in Int\left( G\right)
,\forall w_{2}\in G.
\end{equation*}

Below $C>0$ denotes different constants depending on the same parameters as
ones in Theorem 2.

\subsection{Discussion of Theorem 2}

\label{sec:2.4}

Even though numerical studies are outside of the scope of the current
publications, we briefly discuss in this subsection some computational
aspects of Theorem 2. We point out that it is well known that computations
often work well under conditions which are less restrictive than the theory
is. Still, the theory, such as, e.g. the one of this paper, usually provides
an important guidance for computations.

The maximal value $M$ of the function $\psi \left( x,t\right) $ is achieved
at the point $\left( \overline{x}\left( x_{0}\right) ,0\right) ,$ such that $%
\overline{x}\left( x_{0}\right) \in \partial \Omega $ and $\left\vert 
\overline{x}\left( x_{0}\right) -x_{0}\right\vert ^{2}=\max_{x\in \overline{%
\Omega }}$ $\left\vert x-x_{0}\right\vert ^{2}.$ On the other hand, since
the function $\varphi _{\lambda }^{2}\left( x,t\right) $ changes rapidly
with respect to $\left( x,t\right) ,$ then it seems to be that the
coefficient $c\left( x\right) $ will be accurately reconstructed in
practical computations only in a small neighborhood of the point $\overline{x%
}\left( x_{0}\right) $. Changing $x_{0},$ one can cover a neighborhood of a
part $\partial ^{\prime }\Omega $ of the boundary $\partial \Omega $.

We point out that our \emph{ultimate goal} is to apply the technique of this
paper to our experimental data, which are described in \cite{BTKF}. Next, we
will compare its performance on these data with the performance of the
globally convergent algorithm of \cite{BK}, which was used in \cite{BTKF}. A
potential application of the work \cite{BTKF} is in imaging and
identification of explosive devices. Those devices typically have small
sizes. The data of \cite{BTKF} were collected on a part of a plane in the
backscattering case. The distance between this plane and the front surface
of any target of interest was about 80 centimeters (cm). Using the arrival
time of the backscattering signal, these distances were accurately estimated
for all targets.

The raw data of \cite{BTKF} are very far from the range of the operator of
the forward problem. Hence, it was necessary to use a heuristic data
pre-processing procedure as a preliminary step before applying any
reconstruction algorithm. The pre-processed data were used as the input for
the algorithm of \cite{BK}. One of steps of data pre-processing was data
propagation, which provides an approximation of the data at a part $\Gamma
^{\prime }$ of a plane $\Gamma .$ The distances between $\Gamma ^{\prime }$
and the front surface of a target can vary from 0 cm to 4 cm. Since targets
of interest have rather small sizes of a few centimeters, then imaging of a
small neighborhood of $\Gamma ^{\prime }$ might be sufficient for the
experimental data of \cite{BTKF}. Besides, using a layer stripping
procedure, one might cover a somewhat larger neighborhood of $\Gamma
^{\prime }$.

Next, the domain $\Omega $ was chosen, where the solution of the inverse
problem was computed. The surface $\Gamma ^{\prime }\subset \partial \Omega $
was a part of the boundary of this domain. Another step of the data
pre-processing procedure of \cite{BTKF} was complementing the data on $%
\Gamma ^{\prime }$ by the data on the rest of the boundary $\partial \Omega $%
. Those additional data resulted from the solution of the forward problem
for equation (\ref{2.7}) for the case $c\left( x\right) \equiv 1.$ Accurate
reconstructions were obtained in \cite{BTKF} even for the most difficult
cases of completely blind data. We believe, therefore, that the technique of
the current paper might be applicable to the case of backscattering data, if
complementing those data as in \cite{BTKF}.

In Theorems 3-7 below we assume that the point of minimum of our functional
is an interior point of either the set $G$ or its finite dimensional analog.
For such a point, the condition of the fourth line of (\ref{2.22}) becomes $%
\left( \nabla A\left( v\right) \left( x\right) ,x-x_{0}\right) =\left(
\nabla c\left( x\right) ,x-x_{0}\right) >0.$ Thus, $c\left( x\right) \equiv
const.$ does not satisfy this condition. However, the case $c\left( x\right)
\equiv const.$ as the solution of our CIP is of no interest to us, since the
experimental data of \cite{BTKF} are quite different for this scenario from
the case when a target of interest is present.

As to the condition (\ref{2.104}), in principle it would be better to assume
instead that $f\left( x\right) =\delta \left( x-x^{\prime }\right) $ for a
certain point $x^{\prime }\in \Omega ^{\prime }\diagdown \Omega .$ However,
it is well known that the technique of \cite{BukhK} does not work in this
case. On the other hand, a narrow Gaussian centered at the point $\left\{
x^{\prime }\right\} $ approximates $\delta \left( x-x^{\prime }\right) $ in
the sense of distributions \cite{V}. Thus, it was pointed out on pages 47,
48 of \cite{BK} and on pages 480, 481 of \cite{K} that if the function $%
\delta \left( x-x^{\prime }\right) $ would be replaced with that Gaussian,
then this would be equivalent to $\delta \left( x-x^{\prime }\right) $ from
the Physics standpoint and would provide only an insignificant difference in
the data $s(x,t),p(x,t).$\ We believe that, for the data of \cite{BTKF},
such a replacement can be handled well computationally by the technique of
the current data since this method is stable (Theorems 4,7 below). On the
other hand, both the method of \cite{BukhK} and the technique of this paper
work in the case of this replacement. To ensure (\ref{2.61}), one should
multiply that Gaussian by a function $\chi \in C^{\infty }\left( \mathbb{R}%
^{3}\right) $ such that $\chi \left( x\right) =1$ in $\Omega $, $\chi \left(
x\right) =0$ in $\mathbb{R}^{3}\diagdown \Omega ^{\prime }$ and the integral
of the resulting product over $\Omega ^{\prime }$ would be equal to unity.

Although the topic of differentiation of noisy data is outside of the scope
of this paper, we now briefly comment on it. Functions $\overline{s}\left(
x,t\right) ,\overline{p}\left( x,t\right) $ amount to the second $t-$%
derivative of measured data $s\left( x,t\right) ,p\left( x,t\right) $ in (%
\ref{2.11}), (\ref{2.13}), which naturally contain noise. By (\ref{2.18})
the function $F$ depends on functions $\overline{s},\overline{p}$ and needs
to have more derivatives. Hence, in Theorems 4 and 7 below we consider the
case when the function $F$ is given with an error. It is well known that the
ill-posed problem of stable differentiation of noisy data can be addressed
via a variety of regularization algorithms, see, e.g. \cite{A}.\ We only
mention that our experience of working with experimental data tells us that
such a question can usually be addressed via either one of methods of \cite%
{A} or, again, a proper heuristic data pre-preprocessing procedure. In fact,
differentiation of noisy data is one of elements of the technique of \cite%
{BK}, and this was successfully done for experimental data in Chapter 5 of 
\cite{BK} and in \cite{BTKF}.

Here is an additional consideration about the differentiation. Although we
obtain zero Dirichlet and Neumann boundary conditions in (\ref{2.183}) for
the function $w$ via the introduction of the function $F$ with properties (%
\ref{2.18}), we actually can prove an analog of the main Theorem 2 in the
case of non-zero boundary conditions $\overline{s},\overline{p}$ in (\ref%
{2.17}) for the function $\widetilde{w}$. However, this is inconvenient for
the practical implementation of the gradient method (\ref{2.200}), since in
this case the term $\gamma J_{\lambda ,\alpha }^{\prime }\left( w_{n}\right) 
$ should have zero boundary conditions.\ It is not immediately clear how to
arrange the latter. On the other hand, if working with the finite difference
approximation of the problem (\ref{2.184}), then it is clear how to arrange
zero boundary conditions for this term: one would need to arrange this term
only for interior points of the domain $Q_{T}$. Thus, in this case one would
need to stably calculate only the second $t-$derivatives of noisy data in (%
\ref{2.140}). It is well known that regularization techniques can handle
well the first and second derivatives of noisy data. An extension of results
of this paper to the case of finite differences amounts to a significant
effort, which is outside of our scope here.

\subsection{Global convergence of the gradient method}

\label{sec:2.5}

We now formulate global convergence theorems of the gradient method of the
minimization of the functional $J_{\lambda ,\alpha }.$ Consider an arbitrary
function $w_{1}\in Int\left( G\right) .$ Let $\gamma >0$ be a number. For
brevity we do not indicate the dependence of functions $w_{n}$ on parameters 
$\lambda ,\alpha ,\gamma $. Consider the sequence $\left\{ w_{n}\right\}
_{n=1}^{\infty }$ of the gradient method, 
\begin{equation}
w_{n+1}=w_{n}-\gamma J_{\lambda ,\alpha }^{\prime }\left( w_{n}\right)
,n=1,2,...  \label{2.200}
\end{equation}

\textbf{Theorem 3}. \emph{Let }$\eta $ \emph{and} $T$\emph{\ be numbers of
Theorem 1, }$\lambda _{0}$ \emph{be the number of Theorem 2 and let }$%
Int\left( G\right) \neq \varnothing $\emph{. Choose parameters }$\lambda
\geq \lambda _{0}$\emph{\ and }$\alpha \geq 2C\exp \left( -\lambda N\right)
. $\emph{\ Assume that the functional }$J_{\lambda ,\alpha }$\emph{\
achieves its minimal value on the set }$G$\emph{\ at a point }$w_{\min }\in
Int\left( G\right) $\emph{.}\ \emph{\ Then such a point }$w_{\min }$ \emph{%
is unique. Consider the sequence (\ref{2.200}), where }$w_{1}\in Int\left(
G\right) $ \emph{is an arbitrary point}. \emph{Assume that }$\left\{
w_{n}\right\} _{n=1}^{\infty }\subset Int\left( G\right) .$ \emph{Then there
exists a sufficiently small number }$\gamma =\gamma \left( \lambda ,\alpha
,\eta ,T,b,R,G\right) \in \left( 0,1\right) $ \emph{and a number }$q=q\left(
\gamma \right) \in \left( 0,1\right) ,$\emph{\ both dependent only on listed
parameters, such that the sequence (\ref{2.200}) converges to the point }$%
w_{\min },$%
\begin{equation}
\left\Vert w_{n+1}-w_{\min }\right\Vert _{H^{6}\left( Q_{T}\right) }\leq
q^{n}\left\Vert w_{1}-w_{\min }\right\Vert _{H^{6}\left( Q_{T}\right) }.
\label{2.2000}
\end{equation}

Following subsection 2.4, assume now that the function $F$ is given with an
error. Then the natural question is on how far are points $w_{\min }$ and $%
w_{n}$ from the exact solution of the problem (\ref{2.184}) with errorless
data. Theorem 4 addresses this question in terms of the norm of the space $%
H^{1}\left( P_{d}\right) .$ The latter might be sufficient for computations.
We use the Tikhonov concept for ill-posed problems mentioned in
Introduction. Namely, we assume that there exists the exact solution $%
w^{\ast }$ of the problem (\ref{2.184}) with errorless data $F^{\ast }\in
H^{6}\left( Q_{T}\right) ,$ 
\begin{equation}
w^{\ast }\in G\left( Q_{T},b,R,f,F^{\ast }\right) :=G^{\ast }\neq
\varnothing .  \label{2.250}
\end{equation}

\textbf{Theorem 4.} \emph{Let }$\eta $ \emph{and} $T$\emph{\ be numbers of
Theorem 1, }$\lambda _{0}$ \emph{be the number of Theorem 2 and let }$%
Int\left( G\right) \neq \varnothing $\emph{. Consider the problem (\ref%
{2.184}). Assume that conditions of Theorem 3 about functions }$%
w_{1},w_{\min }$\emph{\ and} \emph{the sequence (\ref{2.200}) hold and that (%
\ref{2.250}) is valid.} \emph{In addition, assume that the function }$F$ 
\emph{is given with an error of the level }$\delta >0,$\emph{\ i.e. }$%
\left\Vert F-F^{\ast }\right\Vert _{C^{3}\left( \overline{Q}_{T}\right)
}\leq \delta ,$ \emph{where\ }$\delta \in \left( 0,\delta _{0}\right) $\emph{%
, where\ the number }$\delta _{0}\in \left( 0,1\right) $\emph{\ is so small
that} $\delta _{0}\leq \min \left( C_{2}R,\left( 1+b\right) ^{-1}\xi
/2\right) $ \emph{and also} $\lambda _{1}=\ln \left( \delta _{0}^{-1/\left(
2M\right) }\right) \geq \lambda _{0}.$\emph{\ Choose }$\lambda =\lambda
\left( \delta \right) =\ln \left( \delta ^{-1/\left( 2M\right) }\right) $%
\emph{\ and }$\alpha =\alpha \left( \delta \right) =2C\delta ^{N/\left(
2M\right) }$\emph{. Then for the same numbers }$\gamma ,q$ \emph{as in
Theorem 3}%
\begin{equation}
\left\Vert w^{\ast }-w_{\min }\right\Vert _{H^{1}\left( P_{d}\right) }\leq
C\delta ^{\rho },\rho =\min \left( \frac{1}{2},\frac{N}{4M}\right) ,
\label{2.2001}
\end{equation}%
\emph{\ }%
\begin{equation}
\left\Vert w_{n+1}-w^{\ast }\right\Vert _{H^{1}\left( P_{d}\right) }\leq
q^{n}\left\Vert w_{1}-w^{\ast }\right\Vert _{H^{6}\left( Q_{T}\right)
}+C\delta ^{\rho },n=1,2,...  \label{2.2002}
\end{equation}%
\emph{\ Estimates (\ref{2.2001}) and (\ref{2.2002}) remain true in the case
of errorless data with }$\delta =0$ \emph{if setting }$\alpha =\alpha \left(
\lambda \right) =2C\exp \left( -\lambda N\right) $\emph{\ and replacing }$%
C\delta ^{\rho }$\emph{\ with }$C\exp \left( -\lambda N/2\right) $ \emph{for}
$\lambda \geq \lambda _{0}.$

\subsection{ The finite dimensional case}

\label{sec:2.6}

Consider the weighted space $L_{2}^{\lambda }\left( Q_{T}\right) ,$%
\begin{equation*}
L_{2}^{\lambda }\left( Q_{T}\right) =\left\{ w:\left\Vert w\right\Vert
_{L_{2}^{\lambda }\left( Q_{T}\right) }=\left(
\int\limits_{Q_{T}}w^{2}\varphi _{\lambda }dxdt\right) ^{1/2}<\infty
\right\} .
\end{equation*}%
Recall that $Y\left( w\right) $ is the nonlinear operator in the left hand
side of (\ref{2.184}) supplied with initial and boundary conditions (\ref%
{2.183}). Hence, we can consider the operator $Y$ as $Y:G\rightarrow
L_{2}^{\lambda }\left( Q_{T}\right) .$ Let $Y^{\prime }\left( w\right) $ be
the Fr\'{e}chet derivative of $Y$ at the point $w$. Then $J_{\lambda ,\alpha
}^{\prime }\left( w\right) =Y^{\prime \ast }\left( w\right) \left( Y\left(
w\right) \right) ,$ where the linear bounded operator $Y^{\prime \ast
}\left( w\right) :L_{2}^{\lambda }\left( Q_{T}\right) \rightarrow
H_{0}^{6}\left( Q_{T}\right) $ is adjoint to the operator $Y^{\prime }\left(
w\right) ,$ see section 8.1 of \cite{BKS}. Since $L_{2}^{\lambda }\left(
Q_{T}\right) \neq H_{0}^{6}\left( Q_{T}\right) ,$ then it is both not easy
and time consuming to calculate the Fr\'{e}chet derivative $J_{\lambda
,\alpha }^{\prime }\left( w_{n}\right) $ in (\ref{2.200})\emph{\ }for each $%
n $.\ On the other hand, in computations one always works with a finite
dimensional space. In this case one deals with vectors of parameters in an
Euclidian space. Hence, the gradient of a finite dimensional analog of the
functional $J_{\lambda ,\alpha }$ can be easily computed. Thus, analogs of
above theorems for the finite dimensional case might be useful for
computations. These analogs are formulated in the current subsection.

In this subsection we work with finite dimensional subspaces of two spaces: $%
H_{0}^{6}\left( Q_{T}\right) $ and $H_{0}^{3}\left( Q_{T}\right) .$ In the
case of $H_{0}^{6}\left( Q_{T}\right) $ constants in the convergence Theorem
6 for the gradient method are independent on the dimension of the subspace.
Furthermore, Theorem 7 estimates distances between points calculated by the
gradient method and the exact solution $w^{\ast }$. The space $%
H_{0}^{3}\left( Q_{T}\right) $ is simpler to work with. On the other hand,
the constant $C_{3}>0$ in the corresponding Theorem 8 depends on the
dimension of the subspace and those distances are not estimated.

For $m=3,6$ let 
\begin{equation*}
H_{0}^{m}\left( \Omega \right) =\left\{ u\in H^{m}\left( \Omega \right)
:u\mid _{\partial \Omega }=\partial _{\nu }u\mid _{\partial \Omega
}=0\right\} ,H_{0}^{m}\left( 0,T\right) =\left\{ \psi \left( t\right) \in
H^{m}\left( 0,T\right) :\psi ^{\prime }\left( 0\right) =0\right\} .
\end{equation*}%
Let $\left\{ \phi _{i,m}\left( x\right) \right\} _{i=1}^{\infty }\subset
H_{0}^{m}\left( \Omega \right) $ be an orthonormal basis in $H_{0}^{m}\left(
\Omega \right) $ and $\left\{ \psi _{j,m}\left( t\right) \right\}
_{j=1}^{\infty }\subset H_{0}^{m}\left( 0,T\right) $ be an orthonormal basis
in $H_{0}^{m}\left( 0,T\right) .$ For an integer $k\geq 1,$ let $%
H_{0}^{m,k}\left( Q_{T}\right) $ be the subspace of $H_{0}^{m}\left(
Q_{T}\right) $ with the orthonormal basis $\left\{ \phi _{i,m}\left(
x\right) \psi _{j,m}\left( t\right) \right\} _{i,j=1}^{k}.$ Hence, $\dim
\left( H_{0}^{m,k}\left( Q_{T}\right) \right) =k^{2}$ and 
\begin{equation}
v\left( x,t\right) =\sum\limits_{i,j=1}^{k}b_{i,j}\phi _{i,m}\left( x\right)
\psi _{j,m}\left( t\right) ,\left( x,t\right) \in Q_{T},\forall v\in
H_{0}^{m,k}\left( Q_{T}\right) ,  \label{2.202}
\end{equation}%
where $b_{i,j}=b_{i,j}\left( v\right) $ are numbers. Denote $B=\left(
b_{1,1},...,b_{k,k}\right) ^{T}\in \mathbb{R}^{k^{2}}.$ For each vector $%
B\in \mathbb{R}^{k^{2}},$ let $v_{B}\left( x,t\right) $ be the function $%
v\left( x,t\right) $ represented via (\ref{2.202}).

Let $Z_{k}^{m}:H_{0}^{m}\left( Q_{T}\right) \rightarrow H_{0}^{m,k}\left(
Q_{T}\right) $ be the operator of the orthogonal projection of the space $%
H_{0}^{m}\left( Q_{T}\right) $ onto its subspace $H_{0}^{m,k}\left(
Q_{T}\right) .$ Based on (\ref{2.22}), we define the set $G_{m,k}\left(
Q_{T},b,R,f,F\right) :=G_{m,k}\subset \mathbb{R}^{k^{2}}$ as 
\begin{equation*}
G_{m,k}=\left\{ B\right\} \in \mathbb{R}^{k^{2}}:\left\{ 
\begin{array}{c}
\left\vert B\right\vert \leq R, \\ 
v_{B}\in H_{0}^{m,k}\left( Q_{T}\right) , \\ 
\left( 1+b\right) ^{-1}\left( \Delta f\right) \left( x\right) \leq
v_{B}\left( x,0\right) +F\left( x,0\right) \leq \left( \Delta f\right)
\left( x\right) ,\forall x\in \overline{\Omega }, \\ 
\left( \nabla A\left( v_{B}\right) \left( x\right) ,x-x_{0}\right) \geq
0,\forall x\in \overline{\Omega }.%
\end{array}%
\right.
\end{equation*}%
Denote $Int\left( G_{m,k}\right) $ the open set of interior points of the
set $G_{m,k}.$ The convexity of the set $G_{m,k}$ can be proven similarly
with Proposition 1.

\subsubsection{The case of $H_{0}^{6,k}\left( Q_{T}\right) $}

\label{sec:2.6.1}

As in (\ref{2.250}), let $w^{\ast }\in G^{\ast }$ be the exact solution of
the problem (\ref{2.184}) with $F:=F^{\ast }$. Let $w_{k}^{\ast
}=Z_{k}^{6}\left( w^{\ast }\right) .$ Then there exists a function $\theta
\left( k\right) >0$ such that 
\begin{equation}
\left\Vert w^{\ast }-w_{k}^{\ast }\right\Vert _{H^{6}\left( Q_{T}\right)
}\leq \theta \left( k\right) ,\lim_{k\rightarrow \infty }\theta \left(
k\right) =0.  \label{2.205}
\end{equation}%
Consider the following analog $J_{\lambda ,\alpha ,k}:G_{6,k}\rightarrow R$
of the functional $J_{\lambda ,\alpha }$ 
\begin{equation*}
J_{\lambda ,\alpha ,k}\left( B\right) =\int\limits_{Q_{T}}\left[ A\left(
w_{B}\right) \left( w_{B}+F\right) _{tt}-\Delta \left( w_{B}+F\right) \right]
^{2}\varphi _{\lambda }^{2}dxdt+\alpha \left\Vert w\right\Vert _{H^{6}\left(
Q_{T}\right) }^{2},B\in G_{6,k}.
\end{equation*}

\textbf{Theorem 5. }\emph{Let }$\eta $\emph{\ and }$T$\emph{\ be numbers of
Theorem 1, }$\lambda _{0}$ \emph{be the number of Theorem 2 and let }$%
Int\left( G_{6.k}\right) \neq \varnothing $\emph{. Then for all }$\lambda
\geq \lambda _{0},\alpha \geq 2C\exp \left( -\lambda N\right) $ \emph{the
functional }$J_{\lambda ,\alpha ,k}\left( B\right) $\emph{\ is strongly
convex on the set }$G_{6,k}$\emph{\ and estimate (\ref{2.24}) holds for }$%
J_{\lambda ,\alpha ,k}\left( B\right) ,$ \emph{where }$%
w_{1}:=w_{B_{1}},w_{2}:=w_{B_{2}}$\emph{,}$\forall B_{1}\in Int\left(
G_{6,k}\right) ,\forall B_{2}\in G_{6,k}.$

Let $\sigma >0$ be a number which will be chosen in Theorem 6. Consider and
arbitrary point $B_{1}\in Int\left( G_{6.k}\right) $ and define the gradient
method for the functional $J_{\lambda ,\alpha ,k}$ as%
\begin{equation}
B_{n+1}=B_{n}-\sigma \nabla J_{\lambda ,\alpha ,k}\left( B_{n}\right)
,n=1,2,...  \label{2.209}
\end{equation}%
Theorem 6 assures the global convergence of the gradient method (\ref{2.209}%
).

\textbf{Theorem 6.} \emph{Let }$\eta $ \emph{and }$T$\emph{\ be numbers of
Theorem 1, }$\lambda _{0}$ \emph{be the number of Theorem 2 and let }$%
Int\left( G_{6.k}\right) \neq \varnothing $\emph{. Let }$\lambda \geq
\lambda _{0},\alpha \geq 2C\exp \left( -\lambda N\right) $\emph{. Assume
that the functional }$J_{\lambda ,\alpha ,k}$\emph{\ achieves its minimal
value on the set }$G_{6,k}$\emph{\ at a point }$B_{\min }\in Int\left(
G_{6.k}\right) $.\emph{\ Then the point }$B_{\min }$ \emph{is unique.} \emph{%
Assume that in the sequence (\ref{2.209}) }$\left\{ B_{n}\right\}
_{n=1}^{\infty }\subset Int\left( G_{6.k}\right) .$\emph{\ Then there exists
a sufficiently small number }$\sigma =\sigma \left( \lambda ,\alpha
,T,b,R,G_{6,k}\right) \in \left( 0,1\right) $\emph{\ and a number }$%
q=q\left( \sigma \right) \in \left( 0,1\right) ,$\emph{\ both dependent on
only listed parameters but independent on }$k$\emph{, such that the sequence
(\ref{2.209}) converges to the point }$B_{\min },$%
\begin{equation}
\left\vert B_{n+1}-B_{\min }\right\vert \leq q^{n}\left\vert B_{1}-B_{\min
}\right\vert .  \label{2.210}
\end{equation}

We now consider the case of an error in the data (subsection 2.4) and
estimate distances between functions\emph{\ }$w_{n},w_{k,\min }$ and the
exact solution $w^{\ast }$.

\textbf{Theorem 7}. \emph{Let }$\eta $ and $T$\emph{\ be numbers of Theorem
1, }$\lambda _{0}$ \emph{be the number of Theorem 2, let }$Int\left(
G_{6.k}\right) \neq \varnothing $\emph{\ and be valid. Assume that the
function }$F$ \emph{is given with an error of the level }$\delta >0,$\emph{\
i.e. }$\left\Vert F-F^{\ast }\right\Vert _{C^{3}\left( \overline{Q}%
_{T}\right) }\leq \delta ,$\emph{\ }$\delta \in \left( 0,\delta _{0}\right) $%
\emph{, where\ the number }$\delta _{0}\in \left( 0,1\right) $\emph{\ is so
small that} $\delta _{0}\leq \min \left( C_{2}R,\left( 1+b\right) ^{-1}\xi
/2\right) $ \emph{and also} $\lambda _{1}=\ln \left( \delta _{0}^{-1/\left(
2M\right) }\right) \geq \lambda _{0}.$\emph{\ Choose }$\lambda =\lambda
\left( \delta \right) =\ln \left( \delta ^{-1/\left( 2M\right) }\right) $%
\emph{\ and }$\alpha =\alpha \left( \delta \right) =2C\delta ^{N/\left(
2M\right) }$\emph{. Also, let the dimension }$k^{2}=k^{2}\left( \delta
\right) $\emph{\ of the space }$H_{0}^{6,k}\left( Q_{T}\right) $\emph{\ be
so large that }$\theta \left( k\left( \delta \right) \right) \leq \delta ,$ 
\emph{where the function }$\theta \left( k\right) $\emph{\ is defined in (%
\ref{2.205}). Assume that conditions of Theorem 6 about }$B_{1},B_{\min }$ 
\emph{and the sequence (\ref{2.209}) hold.\ Denote }$w_{k,\min }=w_{B_{\min
}}$\emph{. Then for the same numbers }$\sigma $\emph{\ and }$q$\emph{\ as in
Theorem 6}%
\begin{equation}
\left\Vert w_{k,\min }-w^{\ast }\right\Vert _{H^{1}\left( P_{d}\right) }\leq
C\delta ^{\rho },\rho =\min \left( \frac{1}{2},\frac{N}{4M}\right) ,
\label{100}
\end{equation}%
\emph{\ }%
\begin{equation}
\left\Vert w_{B_{n+1}}-w^{\ast }\right\Vert _{H^{1}\left( P_{d}\right) }\leq
q^{n}\left\vert B_{1}-B_{\min }\right\vert +C\delta ^{\rho },n=1,2,...
\label{2.211}
\end{equation}%
\emph{In the case of errorless data with }$\delta =0$\emph{\ we set }$%
\lambda =\lambda \left( k\right) =\ln \left( \theta \left( k\right)
^{-1/\left( 2M\right) }\right) \geq \lambda _{0}$\emph{, }$\alpha =\alpha
\left( k\right) =2C\theta ^{\rho }\left( k\right) $ \emph{for} \emph{%
sufficiently large }$k.$\emph{\ Then estimates (\ref{100}) and (\ref{2.211})
are valid, where }$\delta ^{\rho }$\emph{\ is replaced with }$\theta ^{\rho
}\left( k\right) $\emph{.}

\subsubsection{The case of $H_{0}^{3,k}\left( Q_{T}\right) $}

\label{sec:2.6.2}

Consider the orthonormal basis $\left\{ \phi _{i,3}\left( x\right) \psi
_{j,3}\left( t\right) \right\} _{i,j=1}^{k}$ in $H_{0}^{3,k}\left(
Q_{T}\right) $, see (\ref{2.202}). Define the functional $\overline{J}%
_{\lambda }:G_{3,k}\rightarrow R,$ 
\begin{equation}
\overline{J}_{\lambda }\left( B\right) =\int\limits_{Q_{T}}\left[ A\left(
w_{B}\right) \left( w_{B}+F\right) _{tt}-\Delta \left( w_{B}+F\right) \right]
^{2}\varphi _{\lambda }^{2}dxdt,B\in G_{3,k}.  \label{2.212}
\end{equation}%
Theorem 8 for the functional $\overline{J}_{\lambda }$ is an analog of
Theorem 2. Unlike Theorems 5-7, the constant $C_{3}$ of Theorem 8 depends on 
$k$. On the other hand, allowing this dependence, enables us not to include
the regularization term with $\alpha $ in (\ref{2.212}).

\textbf{Theorem 8}. \emph{Let }$\eta $ \emph{and} $T$\emph{\ be numbers of
Theorem 1 and let }$Int\left( G_{3,k}\right) \neq \varnothing $\emph{.
Assume that for }$i,j=1,...,k$\emph{\ functions }$D^{\beta }\phi
_{i,3}\left( x\right) \in L_{\infty }\left( \Omega \right) ,\psi
_{j,3}^{\left( s\right) }\left( t\right) \in L_{\infty }\left( 0,T\right) ,$%
\emph{\ where }$\left\vert \beta \right\vert \leq 3$\emph{\ and }$s=0,1,2,3$%
\emph{.\ Then there exists a sufficiently large number }$\lambda
_{2}=\lambda _{2}\left( \eta ,T,b,R,G_{3,k},k\right) >1$\emph{\ depending
only on listed parameters, such that for all }$\lambda \geq \lambda _{2}$%
\emph{\ the functional }$\overline{J}_{\lambda }\left( B\right) $\emph{\ is
strongly convex on the set }$G_{3,k}.$\emph{\ More precisely, there exists a
constant }$C_{3}=C_{3}\left( \eta ,T,b,R,G_{3,k},P_{d},k\right) >0$\emph{\
such that\ }%
\begin{equation}
\overline{J}_{\lambda }\left( B_{2}\right) -\overline{J}_{\lambda }\left(
B_{1}\right) -\left( \nabla \overline{J}_{\lambda }^{\prime }\left(
B_{1}\right) ,B_{2}-B_{1}\right) _{k}\geq C_{3}\exp \left( \lambda d\right)
\left\vert B_{2}-B_{1}\right\vert ^{2},  \label{2.214}
\end{equation}%
$\forall B_{1}\in Int\left( G_{3,k}\right) ,\forall B_{2}\in G_{3,k},$ \emph{%
where }$\left( ,\right) _{k}$\emph{\ is the scalar product in }$\mathbb{R}%
^{k^{2}}.$ \emph{In addition, the analog of Theorems 6 is valid, in which }$%
\lambda _{0}$ \emph{and }$G_{6,k}$ \emph{are replaced with }$\lambda _{2}$ 
\emph{and }$G_{3,k}$\emph{\ respectively and the parameter }$\alpha $ \emph{%
is not counted.}

\section{Proof of Theorem 1}

\label{sec:3}

In this proof $C_{1}>0$ denotes different positive constants depending on
the same parameters as in the formulation of this theorem. First, we
estimate from the below the term with the function $g$ in (\ref{2.131}). We
have%
\begin{equation}
u\left( x,0\right) =u\left( x,t\right) -\int\limits_{0}^{t}u_{t}\left(
x,\tau \right) d\tau .  \label{3.000}
\end{equation}%
Hence, 
\begin{equation}
g\left( x,t\right) u\left( x,0\right) u_{tt}\left( x,t\right) \varphi
_{\lambda }^{2}=\left( \left( gu_{tt}u\right) \left( x,t\right) -\left(
gu_{tt}\right) \left( x,t\right) \int\limits_{0}^{t}u_{t}\left( x,\tau
\right) d\tau \right) \varphi _{\lambda }^{2}\left( x,t\right) .
\label{3.001}
\end{equation}%
Since $\varphi _{\lambda }^{2}\left( x,t\right) =\exp \left[ 2\lambda \left(
\left\vert x-x_{0}\right\vert ^{2}-\eta t^{2}\right) \right] ,$ then,
applying the Cauchy inequality, we obtain 
\begin{equation*}
\left( gu_{tt}u\right) \left( x,t\right) \varphi _{\lambda }^{2}\left(
x,t\right) =\partial _{t}\left[ gu_{t}u\varphi _{\lambda }^{2}\right]
-gu_{t}^{2}\varphi _{\lambda }^{2}-g_{t}u_{t}u\varphi _{\lambda
}^{2}+4\lambda \eta tgu_{t}u\varphi _{\lambda }^{2}
\end{equation*}%
\begin{equation}
\geq \partial _{t}\left[ gu_{t}u\varphi _{\lambda }^{2}\right]
-C_{1}u_{t}^{2}\varphi _{\lambda }^{2}-C_{1}\lambda ^{2}u^{2}\varphi
_{\lambda }^{2}.  \label{3.002}
\end{equation}%
Now we estimate the term with the integral in (\ref{3.001}),%
\begin{equation*}
-\varphi _{\lambda }^{2}\left( gu_{tt}\right) \left( x,t\right)
\int\limits_{0}^{t}u_{t}\left( x,\tau \right) d\tau
\end{equation*}%
\begin{equation*}
=\partial _{t}\left( -gu_{t}\varphi _{\lambda
}^{2}\int\limits_{0}^{t}u_{t}\left( x,\tau \right) d\tau \right)
+gu_{t}^{2}\varphi _{\lambda }^{2}+g_{t}u_{t}\varphi _{\lambda
}^{2}\int\limits_{0}^{t}u_{t}\left( x,\tau \right) d\tau -4\lambda \eta
tgu_{t}\varphi _{\lambda }^{2}\int\limits_{0}^{t}u_{t}\left( x,\tau \right)
d\tau
\end{equation*}%
\begin{equation}
\geq \partial _{t}\left[ -gu_{t}\varphi _{\lambda
}^{2}\int\limits_{0}^{t}u_{t}\left( x,\tau \right) d\tau \right]
-C_{1}u_{t}^{2}\varphi _{\lambda }^{2}-C_{1}\left(
\int\limits_{0}^{t}u_{t}\left( x,\tau \right) d\tau \right) ^{2}\varphi
_{\lambda }^{2}  \label{3.003}
\end{equation}%
\begin{equation*}
-C_{1}\lambda \beta u_{t}^{2}\varphi _{\lambda }^{2}-\frac{C_{1}}{\beta }%
\lambda \left( \int\limits_{0}^{t}u_{t}\left( x,\tau \right) d\tau \right)
^{2}\varphi _{\lambda }^{2}
\end{equation*}%
\begin{equation*}
\geq \partial _{t}\left[ -gu_{t}\varphi _{\lambda
}^{2}\int\limits_{0}^{t}u_{t}\left( x,\tau \right) d\tau \right]
-C_{1}\lambda \beta \left( 1+\frac{1}{\lambda \beta }\right)
u_{t}^{2}\varphi _{\lambda }^{2}
\end{equation*}%
\begin{equation*}
-\frac{C_{1}}{\beta }\lambda \left( 1+\frac{\beta }{\lambda }\right) \left(
\int\limits_{0}^{t}u_{t}\left( x,\tau \right) d\tau \right) ^{2}\varphi
_{\lambda }^{2},
\end{equation*}%
where $\beta >0$ is a number which will be chosen later, and it is
independent on $\lambda .$ Here we have used the so-called \textquotedblleft
Cauchy inequality with the parameter $\beta ",$ i.e. $2ab\geq -\beta
a^{2}-b^{2}/\beta ,\forall a,b\in \mathbb{R},\forall \beta >0.$ Summing up (%
\ref{3.002}) and (\ref{3.003}), comparing this sum with (\ref{3.001}) and
taking $\lambda >\max \left( \beta ,1/\beta \right) $, we obtain%
\begin{equation}
gu\left( x,0\right) u_{tt}\varphi _{\lambda }^{2}\geq \partial _{t}\left(
gu_{t}u\varphi _{\lambda }^{2}-gu_{t}\varphi _{\lambda
}^{2}\int\limits_{0}^{t}u_{t}\left( x,\tau \right) d\tau \right)
\label{3.004}
\end{equation}%
\begin{equation*}
-C_{1}\lambda \beta u_{t}^{2}\varphi _{\lambda }^{2}-C_{1}\lambda
^{2}u^{2}\varphi _{\lambda }^{2}-\frac{C_{1}}{\beta }\lambda \left(
\int\limits_{0}^{t}u_{t}\left( x,\tau \right) d\tau \right) ^{2}\varphi
_{\lambda }^{2}.
\end{equation*}%
Integrating (\ref{3.004}) over $Q_{T}$ and recalling that $u_{t}\left(
x,0\right) =0,$ we obtain 
\begin{equation*}
\int\limits_{Q_{T}}gu\left( x,0\right) u_{tt}\varphi _{\lambda }^{2}dxdt\geq
\int\limits_{\Omega _{T}}gu_{t}u\varphi _{\lambda
}^{2}dx-\int\limits_{\Omega _{T}}\left( gu_{t}\int\limits_{0}^{T}u_{t}\left(
x,\tau \right) d\tau \right) \varphi _{\lambda }^{2}dx
\end{equation*}%
\begin{equation}
-C_{1}\int\limits_{Q_{T}}\left( \lambda \beta u_{t}^{2}+\lambda
^{2}u^{2}\right) \varphi _{\lambda }^{2}dxdt-\frac{C_{1}}{\beta }\lambda
\int\limits_{Q_{T}}\left( \int\limits_{0}^{t}u_{t}\left( x,\tau \right)
d\tau \right) ^{2}\varphi _{\lambda }^{2}dxdt.  \label{3.006}
\end{equation}%
Since $\varphi _{\lambda }^{2}\mid _{\Omega _{T}}\leq \exp \left( -2\lambda
N\right) ,$ then, applying the Cauchy inequality, we obtain 
\begin{equation*}
\int\limits_{\Omega _{T}}gu_{t}u\varphi _{\lambda
}^{2}dx-\int\limits_{\Omega _{T}}\left( gu_{t}\int\limits_{0}^{T}u_{t}\left(
x,\tau \right) d\tau \right) \varphi _{\lambda }^{2}dx
\end{equation*}%
\begin{equation}
\geq -C_{1}\exp \left( -2\lambda N\right) \int\limits_{\Omega _{T}}\left(
u_{t}^{2}+u^{2}\right) dx-C_{1}\exp \left( -2\lambda N\right)
\int\limits_{\Omega _{T}}\left( \int\limits_{0}^{T}u_{t}\left( x,\tau
\right) d\tau \right) ^{2}dx  \label{3.007}
\end{equation}%
\begin{equation*}
\geq -C_{1}\exp \left( -2\lambda N\right) \int\limits_{\Omega _{T}}\left(
u_{t}^{2}+u^{2}\right) dx-C_{1}\exp \left( -2\lambda N\right)
T\int\limits_{Q_{T}}u_{t}^{2}dxdt.
\end{equation*}%
Next, since $\varphi _{\lambda }^{2}\left( x,t\right) \geq \exp \left(
2\lambda d\right) $ for $\left( x,t\right) \in P_{d},$ 
\begin{equation*}
\exp \left( 2\lambda d\right) >\exp \left( -2\lambda N\right) T\text{ and }%
\exp \left( -2\lambda N\right) T<\exp \left( -\lambda N\right) \text{ for }%
\lambda \geq \overline{\lambda },
\end{equation*}%
then 
\begin{equation}
-C_{1}\exp \left( -2\lambda N\right) T\int\limits_{Q_{T}}u_{t}^{2}dxdt\geq
-C_{1}\int\limits_{P_{d}}u_{t}^{2}\varphi _{\lambda }^{2}dxdt-C_{1}\exp
\left( -\lambda N\right) \int\limits_{Q_{T}\diagdown P_{d}}u_{t}^{2}dxdt
\label{3.008}
\end{equation}%
\begin{equation*}
\geq -C_{1}\int\limits_{Q_{T}}u_{t}^{2}\varphi _{\lambda }^{2}dxdt-C_{1}\exp
\left( -\lambda N\right) \int\limits_{Q_{T}}u_{t}^{2}dxdt.
\end{equation*}%
The inequality of the second line of (\ref{3.008}) follows from the
inequality of the first line, since $P_{d}\subset Q_{T}$ and $Q_{T}\diagdown
P_{d}\subset Q_{T}.$ By lemma 1.10.3 of \cite{BK} the following estimate
holds%
\begin{equation}
\int\limits_{Q_{T}}\left( \int\limits_{0}^{t}u_{t}\left( x,\tau \right)
d\tau \right) ^{2}\varphi _{\lambda }^{2}dxdt\leq \frac{C_{1}}{\lambda }%
\int\limits_{Q_{T}}u_{t}^{2}\varphi _{\lambda }^{2}dxdt.  \label{3.009}
\end{equation}%
Combining (\ref{3.006}), (\ref{3.007}), (\ref{3.008}) and (\ref{3.009}) and
taking $\lambda >1/\beta ^{2}$, we obtain%
\begin{equation}
\int\limits_{Q_{T}}gu\left( x,0\right) u_{tt}\varphi _{\lambda }^{2}dxdt\geq
-C_{1}\int\limits_{Q_{T}}\left( \lambda \beta u_{t}^{2}+\lambda
^{2}u^{2}\right) \varphi _{\lambda }^{2}dxdt  \label{3.0010}
\end{equation}%
\begin{equation*}
-C_{1}\exp \left( -2\lambda N\right) \int\limits_{\Omega _{T}}\left(
u_{t}^{2}+u^{2}\right) dx-C_{1}\exp \left( -\lambda N\right)
\int\limits_{Q_{T}}u_{t}^{2}dxdt.
\end{equation*}%
The Carleman estimate of Theorem 1.10.2 of \cite{BK} for the hyperbolic
operator $c\left( x\right) \partial _{t}^{2}-\Delta $ leads to 
\begin{equation}
\int\limits_{Q_{T}}\left( c\left( x\right) u_{tt}-\Delta u\right)
^{2}\varphi _{\lambda }^{2}dxdt+C_{1}\lambda ^{3}\exp \left( 2\lambda
M\right) \left( \left\Vert \partial _{\nu }u\mid _{S_{T}}\right\Vert
_{L_{2}\left( S_{T}\right) }^{2}+\left\Vert u\right\Vert _{H^{1}\left(
S_{T}\right) }^{2}\right)  \label{3.0011}
\end{equation}%
\begin{equation*}
+C_{1}\lambda ^{3}\exp \left( -2\lambda N\right) \left( \left\Vert
u_{t}\right\Vert _{L_{2}\left( \Omega _{T}\right) }^{2}+\left\Vert
u\right\Vert _{H^{1}\left( \Omega _{T}\right) }^{2}\right) \geq
C_{1}\int\limits_{Q_{T}}\left( \lambda \left( \nabla u\right) ^{2}+\lambda
u_{t}^{2}+\lambda ^{3}u^{2}\right) \varphi _{\lambda }^{2}dxdt.
\end{equation*}%
Sum up (\ref{3.0010}) with (\ref{3.0011}) and choose $\beta =1/2$. Then we
obtain (\ref{2.131}). The assertion of this theorem about $c\left( x\right)
\equiv 1$ follows from the above and Corollary 1.10.2 of \cite{BK}. $\square 
$

\section{Proofs of Theorem 2 and Corollary 1}

\label{sec:4}

\textbf{Proof of Theorem 2. }Consider two arbitrary functions $w_{1}\in
Int\left( G\right) ,w_{2}\in G$. Let $h=w_{2}-w_{1}.$ Since $h\in
H_{0}^{6}\left( Q_{T}\right) ,$ then 
\begin{equation}
h_{t}\left( x,0\right) =0,h\mid _{S_{T}}=\partial _{\nu }h\mid _{S_{T}}=0.
\label{4.1}
\end{equation}%
Next, the triangle inequality combined with the second line of (\ref{2.22})
implies that $\left\Vert h\right\Vert _{H^{6}\left( Q_{T}\right) }\leq 2R.$
Hence, (\ref{2.20}) leads to 
\begin{equation}
\left\Vert h\right\Vert _{C^{3}\left( \overline{Q}_{T}\right) }\leq C.
\label{4.2}
\end{equation}%
First, we evaluate the difference $J_{\lambda ,\alpha }\left( w_{1}+h\right)
-J_{\lambda ,\alpha }\left( w_{1}\right) $ and single out the linear term
with respect to $h$, since this term is $J_{\lambda ,\alpha }^{\prime
}\left( w_{1}\right) \left( h\right) .$ Denote%
\begin{equation}
I_{1}=\left[ A\left( w_{1}+h\right) \left( w_{1}+h+F\right) _{tt}-\Delta
\left( w_{1}+h+F\right) \right] ^{2}.  \label{4.20}
\end{equation}%
Then%
\begin{equation}
J_{\lambda ,\alpha }\left( w_{2}\right) =J_{\lambda ,\alpha }\left(
w_{1}+h\right) =\dint\limits_{Q_{T}}I_{1}\varphi _{\lambda }^{2}dxdt+\alpha
\left\Vert w_{1}+h\right\Vert _{H^{6}\left( Q_{T}\right) }^{2}.
\label{4.201}
\end{equation}%
By (\ref{2.21})%
\begin{equation*}
A\left( w_{1}+h\right)
\end{equation*}%
\begin{equation*}
=\left( \Delta f\right) \left( x\right) \left( \frac{1}{\left(
w_{1}+F\right) \left( x,0\right) }-\frac{h\left( x,0\right) }{\left(
w_{1}+F\right) ^{2}\left( x,0\right) }+\frac{h^{2}\left( x,0\right) }{\left(
w_{1}+F\right) ^{2}\left( x,0\right) \left( w_{1}+F+h\right) \left(
x,0\right) }\right) .
\end{equation*}%
Hence,%
\begin{equation*}
A\left( w_{1}+h\right) =A\left( w_{1}\right) -\frac{A\left( w_{1}\right) }{%
\left( w_{1}+F\right) \left( x,0\right) }h\left( x,0\right) +\frac{A\left(
w_{1}+h\right) }{\left( w_{1}+F\right) ^{2}\left( x,0\right) }h^{2}\left(
x,0\right) .
\end{equation*}%
Hence,%
\begin{equation*}
A\left( w_{1}+h\right) \left( w_{1}+h+F\right) _{tt}=A\left( w_{1}+h\right)
\left( w_{1}+F\right) _{tt}+A\left( w_{1}+h\right) h_{tt}=Q_{1}+Q_{2}.
\end{equation*}%
\begin{equation*}
Q_{1}=A\left( w_{1}+h\right) \left( w_{1}+h+F\right) _{tt}=A\left(
w_{1}\right) \left( w_{1}+F\right) _{tt}
\end{equation*}%
\begin{equation*}
-\frac{A\left( w_{1}\right) }{\left( w_{1}+F\right) \left( x,0\right) }%
\left( w_{1}+F\right) _{tt}h\left( x,0\right) +\frac{A\left( w_{1}+h\right) 
}{\left( w_{1}+F\right) ^{2}\left( x,0\right) }\left( w_{1}+F\right)
_{tt}h^{2}\left( x,0\right) .
\end{equation*}%
\begin{equation*}
Q_{2}=A\left( w_{1}+h\right) h_{tt}=A\left( w_{1}\right) h_{tt}-\frac{%
A\left( w_{1}\right) }{\left( w_{1}+F\right) \left( x,0\right) }h\left(
x,0\right) h_{tt}+\frac{A\left( w_{1}+h\right) }{\left( w_{1}+F\right)
^{2}\left( x,0\right) }h_{tt}h^{2}\left( x,0\right) .
\end{equation*}%
Summing up $Q_{1}$ and $Q_{2},$ we obtain%
\begin{equation*}
A\left( w_{1}+h\right) \left( w_{1}+h+F\right) _{tt}=Q_{1}+Q_{2}
\end{equation*}%
\begin{equation*}
=A\left( w_{1}\right) \left( w_{1}+F\right) _{tt}+A\left( w_{1}\right)
h_{tt}-\frac{A\left( w_{1}\right) }{\left( w_{1}+F\right) \left( x,0\right) }%
\left( w_{1}+F\right) _{tt}h\left( x,0\right)
\end{equation*}%
\begin{equation*}
-\frac{A\left( w_{1}\right) }{\left( w_{1}+F\right) \left( x,0\right) }%
h\left( x,0\right) h_{tt}+\frac{A\left( w_{1}+h\right) }{\left(
w_{1}+F\right) ^{2}\left( x,0\right) }\left[ \left( w_{1}+F\right)
_{tt}+h_{tt}\right] h^{2}\left( x,0\right) .
\end{equation*}%
Hence, by (\ref{4.20})%
\begin{equation*}
I_{1}=
\end{equation*}%
\begin{equation*}
\left\{ \left[ A\left( w_{1}\right) \left( w_{1}+F\right) _{tt}-\Delta
\left( w_{1}+F\right) \right] +\left( A\left( w_{1}\right) h_{tt}-\Delta h-%
\frac{A\left( w_{1}\right) }{\left( w_{1}+F\right) \left( x,0\right) }\left(
w_{1}+F\right) _{tt}h\left( x,0\right) \right) +Z\right\} ^{2},
\end{equation*}%
\begin{equation}
Z=-\frac{A\left( w_{1}\right) }{\left( w_{1}+F\right) \left( x,0\right) }%
h\left( x,0\right) h_{tt}+\frac{A\left( w_{1}+h\right) }{\left(
w_{1}+F\right) ^{2}\left( x,0\right) }\left[ \left( w_{1}+F\right)
_{tt}+h_{tt}\right] h^{2}\left( x,0\right) .  \label{4.3}
\end{equation}%
Hence,%
\begin{equation*}
I_{1}=\left[ A\left( w_{1}\right) \left( w_{1}+F\right) _{tt}-\Delta \left(
w_{1}+F\right) \right] ^{2}
\end{equation*}%
\begin{equation*}
+2\left[ A\left( w_{1}\right) \left( w_{1}+F\right) _{tt}-\Delta \left(
w_{1}+F\right) \right] \left( A\left( w_{1}\right) h_{tt}-\Delta h-\frac{%
A\left( w_{1}\right) }{\left( w_{1}+F\right) \left( x,0\right) }\left(
w_{1}+F\right) _{tt}h\left( x,0\right) \right)
\end{equation*}%
\begin{equation*}
+2\left[ A\left( w_{1}\right) \left( w_{1}+F\right) _{tt}-\Delta \left(
w_{1}+F\right) \right] Z
\end{equation*}%
\begin{equation*}
+\left( A\left( w_{1}\right) h_{tt}-\Delta h-\frac{A\left( w_{1}\right) }{%
\left( w_{1}+F\right) \left( x,0\right) }\left( w_{1}+F\right) _{tt}h\left(
x,0\right) \right) ^{2}
\end{equation*}%
\begin{equation*}
+2\left( A\left( w_{1}\right) h_{tt}-\Delta h-\frac{A\left( w_{1}\right) }{%
\left( w_{1}+F\right) \left( x,0\right) }\left( w_{1}+F\right) _{tt}h\left(
x,0\right) \right) Z+Z^{2}.
\end{equation*}%
Let $I_{1,linear}$ be the part of $I_{1},$ which is linear with respect to $%
h $, 
\begin{equation}
I_{1,linear}=2\left[ A\left( w_{1}\right) \left( w_{1}+F\right) _{tt}-\Delta
\left( w_{1}+F\right) \right]  \label{4.299}
\end{equation}%
\begin{equation*}
\times \left( A\left( w_{1}\right) h_{tt}-\Delta h-\frac{A\left(
w_{1}\right) }{\left( w_{1}+F\right) \left( x,0\right) }\left(
w_{1}+F\right) _{tt}h\left( x,0\right) \right) .
\end{equation*}%
Hence, $I_{1,linear}$ generates the Fr\'{e}chet derivative, 
\begin{equation}
J_{\lambda ,\alpha }^{^{\prime }}\left( w_{1}\right) \left( h\right)
=\dint\limits_{Q_{T}}I_{1,linear}\varphi _{\lambda }^{2}dxdt+2\alpha \left[
w_{1},h\right] ,  \label{4.300}
\end{equation}%
where $\left[ ,\right] $ is the scalar product in $H^{6}\left( P_{d}\right)
. $ Denote 
\begin{equation}
I_{2}=I_{1}-\left[ A\left( w_{1}\right) \left( w_{1}+F\right) _{tt}-\Delta
\left( w_{1}+F\right) \right] ^{2}-I_{1,linear}.  \label{4.30}
\end{equation}%
Ignoring the term with $Z^{2}$ in $I_{2}$ and using the Cauchy inequality,
we obtain 
\begin{equation*}
I_{2}\geq \left( A\left( w_{1}\right) h_{tt}-\Delta h-\frac{A\left(
w_{1}\right) }{\left( w_{1}+F\right) \left( x,0\right) }\left(
w_{1}+F\right) _{tt}h\left( x,0\right) \right) ^{2}
\end{equation*}%
\begin{equation*}
+2\left( A\left( w_{1}\right) h_{tt}-\Delta h-\frac{A\left( w_{1}\right) }{%
\left( w_{1}+F\right) \left( x,0\right) }\left( w_{1}+F\right) _{tt}h\left(
x,0\right) \right) Z
\end{equation*}%
\begin{equation*}
+2\left[ A\left( w_{1}\right) \left( w_{1}+F\right) _{tt}-\Delta \left(
w_{1}+F\right) \right] Z
\end{equation*}%
\begin{equation}
\geq \frac{1}{2}\left( A\left( w_{1}\right) h_{tt}-\Delta h\right)
^{2}-Ch^{2}\left( x,0\right)  \label{4.301}
\end{equation}%
\begin{equation*}
+2\left( A\left( w_{1}\right) h_{tt}-\Delta h-\frac{A\left( w_{1}\right) }{%
\left( w_{1}+F\right) \left( x,0\right) }\left( w_{1}+F\right) _{tt}h\left(
x,0\right) \right) Z
\end{equation*}%
\begin{equation*}
+2\left[ A\left( w_{1}\right) \left( w_{1}+F\right) _{tt}-\Delta \left(
w_{1}+F\right) \right] Z.
\end{equation*}%
We now use the expression (\ref{4.3}) for $Z$,%
\begin{equation*}
2\left( A\left( w_{1}\right) h_{tt}-\Delta h-\frac{A\left( w_{1}\right) }{%
\left( w_{1}+F\right) \left( x,0\right) }\left( w_{1}+F\right) _{tt}h\left(
x,0\right) \right) Z
\end{equation*}%
\begin{equation*}
=-\left[ \frac{2A\left( w_{1}\right) }{\left( w_{1}+F\right) \left(
x,0\right) }\left( A\left( w_{1}\right) h_{tt}-\Delta h-\frac{A\left(
w_{1}\right) }{\left( w_{1}+F\right) \left( x,0\right) }\left(
w_{1}+F\right) _{tt}h\left( x,0\right) \right) \right] h\left( x,0\right)
h_{tt}
\end{equation*}%
\begin{equation*}
+2\left( A\left( w_{1}\right) h_{tt}-\Delta h-\frac{A\left( w_{1}\right) }{%
\left( w_{1}+F\right) \left( x,0\right) }\left( w_{1}+F\right) _{tt}h\left(
x,0\right) \right) \frac{A\left( w_{1}+h\right) }{\left( w_{1}+F\right)
^{2}\left( x,0\right) }\left[ \left( w_{1}+F\right) _{tt}+h_{tt}\right]
\end{equation*}%
\begin{equation*}
\times h^{2}\left( x,0\right) .
\end{equation*}%
Hence,%
\begin{equation}
2\left( A\left( w_{1}\right) h_{tt}-\Delta h-\frac{A\left( w_{1}\right) }{%
\left( w_{1}+F\right) \left( x,0\right) }\left( w_{1}+F\right) _{tt}h\left(
x,0\right) \right) Z  \label{4.302}
\end{equation}%
\begin{equation*}
\geq -\left\{ \frac{2A\left( w_{1}\right) }{\left( w_{1}+F\right) \left(
x,0\right) }\left( A\left( w_{1}\right) h_{tt}-\Delta h-\frac{A\left(
w_{1}\right) }{\left( w_{1}+F\right) \left( x,0\right) }\left(
w_{1}+F\right) _{tt}h\left( x,0\right) \right) \right\} h\left( x,0\right)
h_{tt}
\end{equation*}%
\begin{equation*}
-Ch^{2}\left( x,0\right) ,
\end{equation*}%
Similarly%
\begin{equation*}
2\left[ A\left( w_{1}\right) \left( w_{1}+F\right) _{tt}-\Delta \left(
w_{1}+F\right) \right] Z
\end{equation*}%
\begin{equation}
\geq -\left\{ \frac{2A\left( w_{1}\right) }{\left( w_{1}+F\right) \left(
x,0\right) }\left[ A\left( w_{1}\right) \left( w_{1}+F\right) _{tt}-\Delta
\left( w_{1}+F\right) \right] \right\} h\left( x,0\right)
h_{tt}-Ch^{2}\left( x,0\right) .  \label{4.303}
\end{equation}%
Denote 
\begin{equation*}
g\left( x,t\right) =-\left\{ \frac{2A\left( w_{1}\right) }{\left(
w_{1}+F\right) \left( x,0\right) }\left( A\left( w_{1}\right) h_{tt}-\Delta
h-\frac{A\left( w_{1}\right) }{\left( w_{1}+F\right) \left( x,0\right) }%
\left( w_{1}+F\right) _{tt}h\left( x,0\right) \right) \right\}
\end{equation*}%
\begin{equation}
-\left\{ \frac{2A\left( w_{1}\right) }{\left( w_{1}+F\right) \left(
x,0\right) }\left[ A\left( w_{1}\right) \left( w_{1}+F\right) _{tt}-\Delta
\left( w_{1}+F\right) \right] \right\} .  \label{4.304}
\end{equation}%
It follows from (\ref{2.19}), (\ref{2.22}) and (\ref{4.304}) that functions $%
g,g_{t}\in C\left( \overline{Q}_{T}\right) $ and $\left\Vert g\right\Vert
_{C\left( \overline{Q}_{T}\right) },\left\Vert g_{t}\right\Vert _{C\left( 
\overline{Q}_{T}\right) }\leq C.$ Combining (\ref{3.000}) with (\ref{4.30})-(%
\ref{4.303}), we obtain%
\begin{equation*}
I_{1}-\left[ A\left( w_{1}\right) \left( w_{1}+F\right) _{tt}-\Delta \left(
w_{1}+F\right) \right] ^{2}-I_{1,linear}
\end{equation*}%
\begin{equation*}
\geq \frac{1}{2}\left( A\left( w_{1}\right) h_{tt}-\Delta h\right)
^{2}-Ch^{2}\left( x,t\right) -C\left( \dint\limits_{0}^{t}h_{t}\left( x,\tau
\right) d\tau \right) ^{2}+g\left( x,t\right) h\left( x,0\right) h_{tt}.
\end{equation*}%
Hence, (\ref{2.23}) and (\ref{4.300}) imply that 
\begin{equation*}
J_{\lambda ,\alpha }\left( w_{1}+h\right) -J_{\lambda ,\alpha }\left(
w_{1}\right) -J_{\lambda ,\alpha }^{\prime }\left( w_{1}\right) \left(
h\right)
\end{equation*}%
\begin{equation*}
\geq \frac{1}{2}\int\limits_{Q_{T}}\left( A\left( w_{1}\right) h_{tt}-\Delta
h\right) ^{2}\varphi _{\lambda }^{2}dxdt+\int\limits_{Q_{T}}g\left(
x,t\right) h\left( x,0\right) h_{tt}\varphi _{\lambda }^{2}dxdt
\end{equation*}%
\begin{equation*}
-C\int\limits_{Q_{T}}h^{2}\varphi _{\lambda
}^{2}dxdt-C\int\limits_{Q_{T}}\left( \int\limits_{0}^{t}h_{t}\left( x,\tau
\right) d\tau \right) ^{2}\varphi _{\lambda }^{2}dxdt+\alpha \left\Vert
h\right\Vert _{H^{6}\left( Q_{T}\right) }^{2}.
\end{equation*}%
Applying Theorem 1, (\ref{3.009}) and (\ref{4.1}), we obtain%
\begin{equation}
J_{\lambda ,\alpha }\left( w_{1}+h\right) -J_{\lambda ,\alpha }\left(
w_{1}\right) -J_{\lambda ,\alpha }^{\prime }\left( w_{1}\right) \left(
h\right) \geq C\int\limits_{Q_{T}}\left( \lambda h_{t}^{2}+\lambda \left(
\nabla h\right) ^{2}+\lambda ^{3}h^{2}\right) \varphi _{\lambda }^{2}dxdt
\label{4.8}
\end{equation}

\begin{equation*}
-C\exp \left( -2\lambda N\right) \left( \left\Vert h_{t}\right\Vert
_{L_{2}\left( \Omega _{T}\right) }^{2}+\left\Vert h\right\Vert _{H^{1}\left(
\Omega _{T}\right) }^{2}\right) -C\exp \left( -\lambda N\right) \left\Vert
h_{t}\right\Vert _{L_{2}\left( Q_{T}\right) }^{2}+\alpha \left\Vert
h\right\Vert _{H^{6}\left( Q_{T}\right) }^{2}
\end{equation*}%
Since 
\begin{equation*}
\left\Vert h_{t}\right\Vert _{L_{2}\left( \Omega _{T}\right)
}^{2}+\left\Vert h\right\Vert _{H^{1}\left( \Omega _{T}\right)
}^{2}+\left\Vert h_{t}\right\Vert _{L_{2}\left( Q_{T}\right) }^{2}\leq
C\left\Vert h\right\Vert _{H^{6}\left( Q_{T}\right) }^{2}\text{ and }\alpha
\geq 2C\exp \left( -\lambda N\right) ,
\end{equation*}%
then, using (\ref{4.8}), we obtain%
\begin{equation*}
J_{\lambda ,\alpha }\left( w_{1}+h\right) -J_{\lambda ,\alpha }\left(
w_{1}\right) -J_{\lambda ,\alpha }^{\prime }\left( w_{1}\right) \left(
h\right)
\end{equation*}%
\begin{equation}
\geq C\int\limits_{Q_{T}}\left( \lambda h_{t}^{2}+\lambda \left( \nabla
h\right) ^{2}+\lambda ^{3}h^{2}\right) \varphi _{\lambda }^{2}dxdt+\frac{%
\alpha }{2}\left\Vert h\right\Vert _{H^{6}\left( Q_{T}\right) }^{2},
\label{4.10}
\end{equation}%
$\forall w_{1}\in Int\left( G\right) ,\forall w_{2}=w_{1}+h\in G,$ which is (%
\ref{2.24}). Next, since $P_{d}\subset Q_{T}$ \ and $\varphi _{\lambda
}^{2}\mid _{P_{d}}\geq \exp \left( 2\lambda d\right) ,$ then (\ref{4.10})
implies (\ref{2.242}). $\square $

\textbf{Proof of Corollary 1. }Estimate (\ref{2.243}) follows immediately
from (\ref{2.241}) and (\ref{4.8}). $\square $

\section{Proofs of Theorems 3-8}

\label{sec:5}

\textbf{Proof of Theorem 3. }Recall that $Y\left( w\right) =A\left( w\right)
\left( w+F\right) _{tt}-\Delta \left( w+F\right) $ is the operator in (\ref%
{2.184}) and $Y:G\rightarrow L_{2}^{\lambda }\left( Q_{T}\right) $
(subsection 2.6). The bounded linear operator of the Fr\'{e}chet derivative $%
Y^{\prime }\left( w\right) :H_{0}^{6}\left( Q_{T}\right) \rightarrow
L_{2}^{\lambda }\left( Q_{T}\right) .$ Let $L\left( H_{0}^{6}\left(
Q_{T}\right) ,L_{2}^{\lambda }\left( Q_{T}\right) \right) $ be the space of
bounded linear operators mapping $H_{0}^{6}\left( Q_{T}\right) $ in $%
L_{2}^{\lambda }\left( Q_{T}\right) .$ It follows from results of section
8.2 of \cite{BKS} that if we would prove that the norm of this operator $%
\left\Vert Y^{\prime }\left( w\right) \right\Vert _{L\left( H_{0}^{6}\left(
Q_{T}\right) ,L_{2}^{\lambda }\left( Q_{T}\right) \right) }$ is uniformly
bounded for all $w\in Int\left( G\right) $ and that the map $Y^{\prime
}\left( w\right) :Int\left( G\right) \rightarrow L\left( H_{0}^{6}\left(
Q_{T}\right) ,L_{2}^{\lambda }\left( Q_{T}\right) \right) $ is Lipschitz
continuous on the set $Int\left( G\right) $, then Theorem 2 combined with
conditions of Theorem 3 and the convexity of the set $G$ (Proposition 1)
would imply that the assertion of Theorem 3 is true. It follows from the
above expression (\ref{4.299}) for $I_{1,linear}$ that 
\begin{equation*}
Y^{\prime }\left( w\right) \left( h\right) =A\left( w\right) h_{tt}-\Delta h-%
\frac{A\left( w\right) }{\left( w+F\right) \left( x,0\right) }\left(
w+F\right) _{tt}h\left( x,0\right) ,\forall w\in Int\left( G\right) ,\forall
h\in H_{0}^{6}\left( Q_{T}\right) .
\end{equation*}%
Hence, 
\begin{equation*}
\left\Vert Y^{\prime }\left( w\right) \left( h\right) \right\Vert
_{L_{2}^{\lambda }\left( Q_{T}\right) }\leq C\exp \left( \lambda M\right)
\left\Vert h\right\Vert _{H^{6}\left( Q_{T}\right) },\forall h\in
H_{0}^{6}\left( Q_{T}\right) ,\forall w\in Int\left( G\right) .
\end{equation*}%
Hence, 
\begin{equation}
\left\Vert Y^{\prime }\left( w\right) \right\Vert _{L\left( H_{0}^{6}\left(
Q_{T}\right) ,L_{2}^{\lambda }\left( Q_{T}\right) \right) }\leq C\exp \left(
\lambda M\right) ,\forall w\in Int\left( G\right) .  \label{5.100}
\end{equation}%
To prove the Lipschitz continuity of the map $Y^{\prime }\left( w\right)
:Int\left( G\right) \rightarrow L\left( H_{0}^{6}\left( Q_{T}\right)
,L_{2}^{\lambda }\left( Q_{T}\right) \right) ,$ estimate the norm%
\begin{equation*}
\left\Vert Y^{\prime }\left( w_{1}\right) \left( h\right) -Y^{\prime }\left(
w_{2}\right) \left( h\right) \right\Vert _{L_{2}^{\lambda }\left(
Q_{T}\right) },\forall w_{1},w_{2}\in Int\left( G\right) ,\forall h\in
H_{0}^{6}\left( Q_{T}\right) .
\end{equation*}%
We have for $\left( x,t\right) \in Q_{T}$%
\begin{equation*}
\left\vert Y^{\prime }\left( w_{1}\right) \left( h\right) -Y^{\prime }\left(
w_{2}\right) \left( h\right) \right\vert
\end{equation*}%
\begin{equation*}
=\left\vert \left( A\left( w_{1}\right) -A\left( w_{2}\right) \right)
h_{tt}+\left( \frac{A\left( w_{2}\right) }{\left( w_{2}+F\right) \left(
x,0\right) }\left( w_{2}+F\right) _{tt}-\frac{A\left( w_{1}\right) }{\left(
w_{1}+F\right) \left( x,0\right) }\left( w_{2}+F\right) _{tt}\right) h\left(
x,0\right) \right\vert
\end{equation*}%
\begin{equation*}
\leq C\left( \left\vert w_{1}-w_{2}\right\vert +\left\vert
w_{1tt}-w_{2tt}\right\vert \right) \left( \left\vert h_{tt}\right\vert
+\left\vert h\left( x,0\right) \right\vert \right) \leq C\left\Vert
w_{1}-w_{2}\right\Vert _{H^{6}\left( Q_{T}\right) }\left\Vert h\right\Vert
_{H^{6}\left( Q_{T}\right) }.
\end{equation*}%
Hence, 
\begin{equation}
\left\Vert Y^{\prime }\left( w_{1}\right) -Y^{\prime }\left( w_{2}\right)
\right\Vert _{L\left( H_{0}^{6}\left( Q_{T}\right) ,L_{2}^{\lambda }\left(
Q_{T}\right) \right) }\leq C\exp \left( \lambda M\right) \left\Vert
w_{1}-w_{2}\right\Vert _{H^{6}\left( Q_{T}\right) },\forall w_{1},w_{2}\in
Int\left( G\right) .  \label{5.101}
\end{equation}%
Thus, (\ref{5.100}) and (\ref{5.101}) ensure that the operator $Y^{\prime
}\left( w\right) $ of the Fr\'{e}chet derivative is uniformly bounded and
the map $w\rightarrow Y^{\prime }\left( w\right) \in L\left( H_{0}^{6}\left(
Q_{T}\right) ,L_{2}^{\lambda }\left( Q_{T}\right) \right) $ \ is Lipschitz
continuous on the set $Int\left( G\right) $. $\square $

\textbf{Proof of Theorem 4. }Let the function $w\in Int\left( G\right) .$
Since by (\ref{2.250}) $w^{\ast }\in G^{\ast },$ then estimate (\ref{2.131})
and a slight modification of the proof of Theorem 2 lead to the following
analog of (\ref{2.242}) 
\begin{equation*}
J_{\lambda ,\alpha }\left( w^{\ast }\right) -J_{\lambda ,\alpha }\left(
w\right) -J_{\lambda ,\alpha }^{\prime }\left( w\right) \left( w^{\ast
}-w\right) \geq C\exp \left( 2\lambda d\right) \left\Vert w^{\ast
}-w\right\Vert _{H^{1}\left( P_{d}\right) }^{2}
\end{equation*}%
\begin{equation}
-C\lambda ^{3}\exp \left( -2\lambda N\right) \left( \left\Vert \partial
_{t}\left( w^{\ast }-w\right) \right\Vert _{L_{2}\left( \Omega _{T}\right)
}^{2}+\left\Vert w^{\ast }-w\right\Vert _{H^{1}\left( \Omega _{T}\right)
}^{2}\right)  \label{5.1}
\end{equation}

\begin{equation*}
-C\exp \left( -\lambda N\right) \left\Vert w^{\ast }-w\right\Vert
_{L_{2}\left( Q_{T}\right) }^{2}+\alpha \left\Vert w^{\ast }-w\right\Vert
_{H^{6}\left( Q_{T}\right) }^{2}.
\end{equation*}%
Here the function $F,$ rather than $F^{\ast },$ is involved in $J_{\lambda
,\alpha }\left( w^{\ast }\right) .$ Also, we use the fact that by (\ref%
{2.104}), (\ref{2.20}) and (\ref{2.22}) for all $x\in \overline{\Omega }$ 
\begin{equation*}
w^{\ast }\left( x,0\right) +F\left( x,0\right) =w^{\ast }\left( x,0\right)
+F^{\ast }\left( x,0\right) +\left( F-F^{\ast }\right) \left( x,0\right)
\leq \left( \Delta f\right) \left( x\right) +\delta \leq 2C_{2}R,
\end{equation*}%
\begin{equation*}
w^{\ast }\left( x,0\right) +F\left( x,0\right) =w^{\ast }\left( x,0\right)
+F^{\ast }\left( x,0\right) +\left( F-F^{\ast }\right) \left( x,0\right)
\end{equation*}%
\begin{equation*}
\geq \left( 1+b\right) ^{-1}\left( \Delta f\right) \left( x\right) -\delta
\geq \left( 1+b\right) ^{-1}\xi -\delta \geq \left( 1+b\right) ^{-1}\xi /2.
\end{equation*}%
Using the same arguments as ones in the end of the proof of Theorem 2, we
conclude that terms of (\ref{5.1}) with $\exp \left( -2\lambda N\right) $
and $\exp \left( -\lambda N\right) $ are absorbed by the term $\alpha
\left\Vert w^{\ast }-\overline{w}\right\Vert _{H^{6}\left( Q_{T}\right)
}^{2}/2.$ Hence, we obtain%
\begin{equation}
J_{\lambda ,\alpha }\left( w^{\ast }\right) -J_{\lambda ,\alpha }\left(
w\right) -J_{\lambda ,\alpha }^{\prime }\left( w\right) \left( w^{\ast
}-w\right) \geq C\exp \left( 2\lambda d\right) \left\Vert w^{\ast
}-w\right\Vert _{H^{1}\left( P_{d}\right) }^{2}.  \label{5.2}
\end{equation}

We now estimate $J_{\lambda ,\alpha }\left( w^{\ast }\right) $ from the
above. Let the functional $J_{\lambda ,\alpha }^{\ast }\left( w^{\ast
}\right) $ be obtained from $J_{\lambda ,\alpha }\left( w^{\ast }\right) $
via replacement in (\ref{2.23}) $F$ with $F^{\ast }.$ Since the function $%
w^{\ast }$ is the solution of the problem (\ref{2.184}) with $F:=F^{\ast }$,
then $J_{\lambda ,\alpha }^{\ast }\left( w^{\ast }\right) =\alpha \left\Vert
w^{\ast }\right\Vert _{H^{6}\left( Q_{T}\right) }^{2}.$ Hence, representing $%
F$ in $J_{\lambda ,\alpha }\left( w^{\ast }\right) $ as $F=F^{\ast }+\left(
F-F^{\ast }\right) $ and using $\left\Vert F-F^{\ast }\right\Vert
_{C^{3}\left( \overline{Q}_{T}\right) }\leq \delta ,$ we obtain $J_{\lambda
,\alpha }\left( w^{\ast }\right) \leq C\exp \left( 2\lambda M\right) \delta
^{2}+\alpha R^{2}.$ Hence, using $J_{\lambda ,\alpha }^{\prime }\left(
w_{\min }\right) =0,$ we obtain from (\ref{5.2}) 
\begin{equation}
\left\Vert w^{\ast }-w_{\min }\right\Vert _{H^{1}\left( P_{d}\right)
}^{2}\leq C\delta ^{2}\exp \left( 2\lambda M\right) +\alpha R^{2}.
\label{5.3}
\end{equation}%
Choose $\lambda =\lambda \left( \delta \right) $ and $\alpha =\alpha \left(
\delta \right) $ as in the formulation of this theorem. Then $C\delta
^{2}\exp \left( 2\lambda M\right) +\alpha R^{2}\leq C\left( \delta +\delta
^{N/\left( 2M\right) }\right) \leq C\delta ^{2\rho }.$ Hence, (\ref{5.3})
implies that $\left\Vert w^{\ast }-w_{\min }\right\Vert _{H^{1}\left(
P_{d}\right) }\leq C\delta ^{\rho },$ which is (\ref{2.2001}). By the
triangle inequality $\left\Vert w_{n+1}-w_{\min }\right\Vert _{H^{1}\left(
P_{d}\right) }\geq \left\Vert w_{n+1}-w^{\ast }\right\Vert _{H^{1}\left(
P_{d}\right) }-\left\Vert w^{\ast }-w_{\min }\right\Vert _{H^{1}\left(
P_{d}\right) }.$ Hence, combining (\ref{2.2000}) with (\ref{2.2001}), we
obtain $\left\Vert w_{n+1}-w^{\ast }\right\Vert _{H^{1}\left( P_{d}\right)
}\leq q^{n}\left\Vert w_{1}-w_{\min }\right\Vert _{H^{6}\left( Q_{T}\right)
}+C\delta ^{\rho },$ which is (\ref{2.2002}). Considerations in the case of
errorless data are similar. $\square $

\textbf{Proofs of Theorems 5,6. }Theorems 5 and 6 follow immediately from
Theorems 2 and 3 respectively. $\square $

\textbf{Proof of Theorem 7}. Denote $w_{k,\min }=w_{B_{\min }}.$ As in
Theorem 3, 
\begin{equation}
\left\Vert w_{B_{n+1}}-w_{k,\min }\right\Vert _{H^{1}\left( P_{d}\right)
}\leq \left\Vert w_{B_{n+1}}-w_{k,\min }\right\Vert _{H^{6}\left(
Q_{T}\right) }=\left\vert B_{n+1}-B_{\min }\right\vert \leq q^{n}\left\vert
B_{1}-B_{\min }\right\vert .  \label{5.60}
\end{equation}%
We now prove an analog of estimate (\ref{2.2001}). Recall that $w_{k}^{\ast
}=Z_{k}^{6}\left( w^{\ast }\right) $, i.e. the function $w_{k}^{\ast }$ is
the orthogonal projection of the function $w^{\ast }$ on the $k^{2}-$%
dimensional subspace $H_{0}^{6,k}\left( Q_{T}\right) $ of the space $%
H_{0}^{6}\left( Q_{T}\right) .$ Similarly with (\ref{5.2}) we obtain 
\begin{equation}
J_{\lambda ,\alpha }\left( w_{k}^{\ast }\right) \geq \left\Vert w^{\ast
}-w_{k,\min }\right\Vert _{H^{1}\left( P_{d}\right) }^{2}.  \label{5.7}
\end{equation}%
We now estimate $J_{\lambda ,\alpha }\left( w_{k}^{\ast }\right) $ from the
above. We have 
\begin{equation}
J_{\lambda ,\alpha ,k}\left( w_{k}^{\ast }\right) =\widetilde{J}_{\lambda
,k}\left( w_{k}^{\ast }\right) +\alpha \left\Vert w_{k}^{\ast }\right\Vert
_{H^{6}\left( Q_{T}\right) }^{2},  \label{5.8}
\end{equation}%
\begin{equation}
\widetilde{J}_{\lambda ,k}\left( w_{k}^{\ast }\right) =\int\limits_{Q_{T}}%
\left[ A\left( w_{k}^{\ast }\right) \left( w_{k}^{\ast }+F\right)
_{tt}-\Delta \left( w_{k}^{\ast }+F\right) \right] ^{2}\varphi _{\lambda
}^{2}dxdt.  \label{5.9}
\end{equation}%
It follows from (\ref{2.19}) and (\ref{2.205}) that 
\begin{equation*}
\left\vert \left[ A\left( w_{k}^{\ast }\right) \left( w_{k}^{\ast }+F\right)
_{tt}-\Delta \left( w_{k}^{\ast }+F\right) \right] -\left[ A\left( w^{\ast
}\right) \left( w^{\ast }+F^{\ast }\right) _{tt}-\Delta \left( w^{\ast
}+F^{\ast }\right) \right] \right\vert \leq C\left( \delta +\theta \left(
k\right) \right) \leq C\delta .
\end{equation*}%
Since $A\left( w^{\ast }\right) \left( w^{\ast }+F^{\ast }\right)
_{tt}-\Delta \left( w^{\ast }+F^{\ast }\right) =0,$ then (\ref{5.9}) implies
that $\widetilde{J}_{\lambda ,k}\left( w_{k}^{\ast }\right) \leq C\delta
^{2}\exp \left( 2\lambda M\right) .$ Hence, by (\ref{5.7}) and (\ref{5.8}) 
\begin{equation}
\left\Vert w^{\ast }-w_{k,\min }\right\Vert _{H^{1}\left( P_{d}\right)
}^{2}\leq C\delta ^{2}\exp \left( 2\lambda M\right) +C\alpha .  \label{5.10}
\end{equation}%
Choose $\lambda =\lambda \left( \delta \right) ,\alpha =\alpha \left( \delta
\right) $ as in conditions of this theorem. Then (\ref{5.10}) implies that $%
\left\Vert w^{\ast }-w_{k,\min }\right\Vert _{H^{1}\left( P_{d}\right) }\leq
C\delta ^{\rho },$ which is (\ref{100}). Next, by the triangle inequality%
\begin{equation*}
\left\Vert w_{B_{n+1}}-w_{k,\min }\right\Vert _{H^{1}\left( P_{d}\right)
}\geq \left\Vert w_{B_{n+1}}-w^{\ast }\right\Vert _{H^{1}\left( P_{d}\right)
}-\left\Vert w^{\ast }-w_{k,\min }\right\Vert _{H^{1}\left( P_{d}\right) }
\end{equation*}%
\begin{equation*}
\geq \left\Vert w_{B_{n+1}}-w^{\ast }\right\Vert _{H^{1}\left( P_{d}\right)
}-C\delta ^{\rho }.
\end{equation*}%
Hence, $\left\Vert w_{B_{n+1}}-w^{\ast }\right\Vert _{H^{1}\left(
P_{d}\right) }\leq \left\Vert w_{B_{n+1}}-w_{k,\min }\right\Vert
_{H^{1}\left( P_{d}\right) }+C\delta ^{\rho }.$ Combining this with (\ref%
{5.60}), we obtain the desired estimate (\ref{2.211}). Considerations in the
case of errorless data are similar. $\square $

\textbf{Proof of Theorem 8. }In this proof\textbf{\ }$C_{3}=C_{3}\left( \eta
,b,\xi ,G_{3,k},P_{d},k\right) >0$ denotes different constants depending on
listed parameters. For functions $w_{B}\in G_{3,k}$ the functional $%
\overline{J}_{\lambda }$ in (\ref{2.212}) is the same as the functional $%
\widetilde{J}_{\lambda }$ in (\ref{2.241}) with the only difference that $%
\overline{J}_{\lambda }$ depends on the vector $B,$ whereas $\widetilde{J}%
_{\lambda }$ depends on the function $w$. Orthogonalize functions $\left\{
\phi _{i,m}\left( x\right) \psi _{j,m}\left( t\right) \right\} _{i,j=1}^{k}$
in the space $H^{1}\left( P_{d}\right) .$ Then, using (\ref{2.243}), we
obtain (\ref{2.214}) via 
\begin{equation*}
\overline{J}_{\lambda }\left( B_{2}\right) -\overline{J}_{\lambda }\left(
B_{1}\right) -\left( \nabla \overline{J}_{\lambda }\left( B_{1}\right)
,B_{2}-B_{1}\right)
\end{equation*}%
\begin{equation*}
\geq C_{3}\exp \left( 2\lambda d\right) \left\vert B_{2}-B_{1}\right\vert
^{2}-C_{3}\exp \left( -\lambda N\right) \left\vert B_{2}-B_{1}\right\vert
^{2}
\end{equation*}%
\begin{equation*}
\geq C_{3}\exp \left( \lambda d\right) \left\vert B_{2}-B_{1}\right\vert
^{2},\forall B_{1}\in Int\left( G_{3,k}\right) ,\forall B_{2}\in G_{3,k}.%
\text{ }\square
\end{equation*}

\section{A Finite Element Method for the Reconstruction of the Coefficient $%
c(x)$}

\label{sec:6}

In this section we explain how to compute the coefficient $c(x)$ in (\ref%
{2.151}) explicitly via finite elements, as soon as the solution $w$ of the
problem (\ref{2.184}) is computed via the minimization of the functional (%
\ref{2.212}) by the gradient method. In addition, we outline an algorithm of
computing the minimizer of this functional using finite elements. This might
be useful for computations. The space $H_{0}^{3,k}\left( Q_{T}\right) $ is
used since it is easier to work with finite elements of the third order
rather than with those of the sixth order of $H_{0}^{6,k}\left( Q_{T}\right)
.$ Thus, below $w=w_{B},B\in G_{3,k},$ and we rely in Theorem 8. As basis
functions $\left\{ \phi _{i,3}\left( x\right) \right\} _{i=1}^{k}\subset
H^{3}\left( \Omega \right) ,\left\{ \psi _{j,3}\left( t\right) \right\}
_{j=1}^{k}\subset H_{0}^{3}\left( 0,T\right) $ in (\ref{2.202}), we use
standard finite elements of the third order. They can be orthogonalized in
terms of the space $H^{1}\left( P_{d}\right) ,$ since the proof of Theorem 8
requires this: recall that the constant $C_{3}$ in this theorem depends on $%
k $.

Assume that a minimizer $B_{\min }$ of the functional $\overline{J}_{\lambda
}\left( B\right) $ is an interior point of the set $G_{3,k}$. Consider an
arbitrary point $B_{1}\in Int\left( G_{3,k}\right) $ as the initial guess
and assume that all points obtained by the gradient method for the
functional $\overline{J}_{\lambda }\left( B\right) $ belong to $Int\left(
G_{3,k}\right) .$ Then, applying an analog of Theorem 6, we conclude that (%
\ref{2.210}) is true, i.e. the gradient method results in computing the
minimizer $B_{\min }$ as well as of the corresponding function $w_{B_{\min
}}\left( x,t\right) .$

\subsection{Spaces of finite elements}

\label{subsec:6.1}

When minimizing the functional $\overline{J}_{\lambda }\left( B\right) $ in (%
\ref{2.212}), we search for its stationary point satisfying%
\begin{equation}
\nabla \overline{J}_{\lambda }\left( B\right) =0,B\in Int\left(
G_{3,k}\right) .  \label{6.2}
\end{equation}%
Consider a triangulation of the domain $\Omega $ by non-overlapping
tetrahedral elements $K_{j}\subset \Omega $. These elements form the mesh $%
Ms=\{K_{j}\}_{j=1}^{k_{1}}$, where $k_{1}$ is the total number of elements
in $\Omega $, and $\Omega =\cup _{j=1}^{k_{1}}K_{j}.$ We also introduce the
time discretization $D_{j}$ of the time domain $(0,T)$ into subintervals $%
D_{j}=(t_{j-1},t_{j}]$ of the uniform length $\tau
=t_{j}-t_{j-1},j=1,...k_{2}$. For each element $K_{j}\subset \Omega $ let $%
P_{3}(K_{j})$ be the set of polynomials of the third degree defined on $%
K_{j}.$ Similarly with section 76.4 of \cite{EEJ} and sections 3.1, 3.2 of 
\cite{J}, we define the finite element space $V_{x,h,0}$ as%
\begin{equation}
V_{h,x,0}=\left\{ v\in H^{3}\left( \Omega \right) :v\mid _{K_{j}}\in
P_{3}(K_{j}),j=1,...k_{1};v\mid _{\partial \Omega }=\partial _{\nu }v\mid
_{\partial \Omega }=0\right\} .  \label{6.5}
\end{equation}%
Similarly for each time subinterval $D_{j}$ let $P_{3}(D_{j})$ be the set of
polynomials of the third degree defined on $D_{j}.$ We introduce the finite
element space $V_{h,t,0}$ as 
\begin{equation}
V_{h,t,0}=\left\{ v\in H^{3}\left( 0,T\right) :v\mid _{D_{j}}\in
P_{3}(D_{j}),j=1,...k_{2};v_{t}\mid _{t=0}=0\right\} .  \label{6.51}
\end{equation}%
To formulate the finite element method for (\ref{6.2}) we introduce the
finite element spaces $W_{h,0}$ as%
\begin{equation}
W_{h,0}=V_{h,x,0}\times V_{h,t,0}.  \label{6.511}
\end{equation}%
It follows from (\ref{6.5}), (\ref{6.51}) and (\ref{6.511}) that for a
certain integer $k=k\left( k_{1},k_{2}\right) >0$ 
\begin{equation*}
W_{h,0}\subset H_{0}^{3,k}\left( Q_{T}\right) ,\text{ i.e. }w_{t}\left(
x,0\right) =0,w\mid _{S_{T}}=0,\partial _{\nu }w\mid _{S_{T}}=0,\forall w\in
W_{h,0}.
\end{equation*}%
For brevity and without loss of generality assume that $\dim V_{h,x,0}=\dim
V_{h,t,0}:=k=k_{1}=k_{2}.$Consider linear bases $\left\{ \phi _{i}\left(
x\right) \right\} _{i=1}^{k}$and $\left\{ \psi _{j}\left( t\right) \right\}
_{j=1}^{k}$ in spaces $V_{h,x,0}$ and $V_{h,t,0}$ respectively. Unlike
section 2.6, products $\phi _{i}\left( x\right) \psi _{j}\left( t\right) $
are not orthonormal in $H^{3}\left( Q_{T}\right) .$ Still, they are linearly
independent. Thus, similarly with (\ref{2.202}) 
\begin{equation}
w\left( x,t\right) =w_{B}\left( x,t\right)
=\dsum\limits_{i,j=1}^{k}b_{i,j}\phi _{i}\left( x\right) \psi _{j}\left(
t\right) ,\forall \left( x,t\right) \in Q_{T},\forall w\in W_{h,0},
\label{6.52}
\end{equation}%
\begin{equation}
B=\left\{ b_{1,1},...,b_{k,k}\right\} ^{T}\in \mathbb{R}^{k^{2}}.
\label{6.53}
\end{equation}%
To approximate functions $c(x),$ we use the space of piecewise constant
functions $C_{h}$, 
\begin{equation}
C_{h}:=\{v\in L_{2}(\Omega ):v|_{K_{j}}=v_{j}=const.,j=1,...,k_{1}\}.  \notag
\end{equation}

\subsection{A finite element method for the coefficient $c(x)$}

\label{subsec:6.3}

Let $\left[ ,\right] _{2}$ be the scalar product in $L_{2}\left( \Omega
\right) $\emph{. }To compute the function $c(x)$ we formulate the finite
element method for (\ref{2.151}) as: \emph{Suppose that the function }$%
w\left( x,t\right) =w_{B_{\min }}\left( x,t\right) \in W_{h,0}$\emph{\ is
computed as the solution of the problem (\ref{6.2}). Assuming that the
function }$f\left( x\right) \in H^{7}\left( \mathbb{R}^{3}\right) $\emph{\
is known, approximate the function }$c\left( x\right) \in C_{h}$\emph{\ such
that }%
\begin{equation}
\left[ c\left( x\right) w(x,0),v\right] _{2}=\left[ \Delta f,v\right]
_{2},\forall v\in V_{h}.  \label{6.7}
\end{equation}

We express the function $w(x,0)$ as 
\begin{equation}
w(x,0)=\sum_{i=1}^{k}w_{i}\phi _{i}(x),  \label{6.8}
\end{equation}%
where $w_{i}$ are numbers. Consider an auxiliary vector $\widetilde{c}%
=\left( \widetilde{c}_{1},...,\widetilde{c}_{k}\right) ^{T}$ and assume for
a moment that in (\ref{6.7}) 
\begin{equation}
c\left( x\right) w(x,0)=\sum_{i=1}^{k}\widetilde{c}_{i}w_{i}\phi _{i}(x).
\label{6.81}
\end{equation}%
Substituting (\ref{6.8}) into (\ref{6.7}) and choosing $v(x)=\phi _{j}(x),$
we obtain the following system of linear algebraic equations 
\begin{equation}
\sum_{i=1}^{k}\widetilde{c}_{i}\left[ w_{i}\phi _{i},\phi _{j}\right]
_{2}=\sum_{i=1}^{k}\left[ \Delta f,\phi _{j}\right] _{2},j=1,...,k.
\label{6.9}
\end{equation}%
The system (\ref{6.9}) can be rewritten in the matrix form for the unknown
vector $\widetilde{c}$ and the known vector $w=\left( w_{1},...,w_{k}\right)
^{T}$ as 
\begin{equation}
Q\widetilde{c}=Z.  \label{6.10}
\end{equation}%
In (\ref{6.10}) the matrix $Q$ is the block mass matrix in space and $Z$ is
the load vector.

To obtain an explicit scheme for the computation of the vector $\widetilde{c}
$, we approximate the matrix $Q$ by the lumped mass matrix $Q^{L}$ in space,
i.e., the diagonal approximation is obtained by taking the row sum of $Q$ 
\cite{joly}. Thus, we obtain an explicit formula for the computation of the
vector $\widetilde{c}$: 
\begin{equation}
\widetilde{c}=(Q^{L})^{-1}Z.  \label{6.11}
\end{equation}%
Given $\widetilde{c}$ from (\ref{6.11}), we approximate values of $c(x)$ on
every tetrahedron $K_{j}$ as 
\begin{equation*}
c_{j}:=c|_{K_{j}}\approx \frac{1}{\widetilde{k}_{j}}\sum_{i=1}^{\widetilde{k}%
_{j}}\widetilde{c}_{i},j=1,...k_{1},
\end{equation*}%
where $\widetilde{k}_{j}$ is the number of tetrahedra $K_{i}\in Ms$ which
have at least one point of intersection with the boundary of the tetrahedron 
$K_{j}.$ Hence, $K_{j}$ is among them. As to the numbers $\widetilde{c}_{i},$
we define them as components of the vector $\widetilde{c},$ which correspond
to such functions $\phi _{i}(x)$ in (\ref{6.81}), which are third order
polynomials in those tetrahedra $K_{i}$. Thus, so defined vector $\overline{c%
}=\left( c_{1},...,c_{k}\right) ^{T}$ represents a piecewise constant
function $\overline{c}\left( x\right) \in C_{h},$ which we consider as an
approximation for our target unknown coefficient $c\left( x\right) $. The
lumping procedure does not include approximation errors in the case of
linear Lagrange elements. For the case of higher order finite elements
elements we refer to \cite{joly} for an approximation of the mass matrix by
lumped mass matrix.

\subsection{An outline of the algorithm}

\label{subsec:6.4}

We now outline the algorithm of the minimization of the functional $%
\overline{J}_{\lambda }\left( B\right) $ using the gradient method. The
condition (\ref{6.2}) for the minimizer is 
\begin{equation}
\nabla \overline{J}_{\lambda }(B_{\min })=\int\limits_{Q_{T}}\nabla \left[
A\left( w_{B_{\min }}\right) \left( w_{B_{\min }}+F\right) _{tt}-\Delta
\left( w_{B_{\min }}+F\right) \right] ^{2}\varphi _{\lambda
}^{2}dxdt=0,B_{\min }\in Int\left( G_{3,k}\right) \cap W_{h,0}.  \label{6.12}
\end{equation}%
Consider the vector $g^{n}\in \mathbb{R}^{k^{2}}$ defined as 
\begin{equation}
g^{n}=\int\limits_{Q_{T}}\nabla \left[ A\left( w_{B_{n}}\right) \left(
w_{B_{n}}+F\right) _{tt}-\Delta \left( w_{B_{n}}+F\right) \right]
^{2}\varphi _{\lambda }^{2}dxdt,w_{B_{n}}\in W_{h,0},  \label{6.13}
\end{equation}%
where $B_{n}$ is the vector $B$ obtained at $n$ iterations of the gradient
method.

\textbf{Algorithm}

\begin{itemize}
\item[Step 0.] Choose a point $B_{1}\in Int\left( G_{3,k}\right) \cap
W_{h,0} $ and the corresponding function $w_{1},$ which is computed via (\ref%
{6.52}).

\item[Step 1.] Compute the vector $g^{n}$ via (\ref{6.13}).

\item[Step 2.] Update the vector $B_{n}$ in the gradient method similarly
with (\ref{2.209}) as $B_{n+1}=B_{n}-\sigma g^{n}$ where $\sigma $ is
step-size in the gradient method. Also, obtain the corresponding function $%
w_{B_{n+1}}$ via (\ref{6.52}).

\item[Step 3.] Stop computing vectors $B_{n}$ if $\left\vert
g^{n}\right\vert \leq \theta $. Otherwise set $n:=n+1$ and go to step 1.
Here $\theta $ is the tolerance in the gradient method.

\item[Step 4.] Compute the function $c_{h}\in C_{h}$ via the finite element
discretization as in (\ref{6.11}).
\end{itemize}

\begin{center}
\textbf{Acknowledgments}
\end{center}

This research was supported by US Army Research Laboratory and US Army
Research Office grant number W911NF-11-1-0399, the Swedish Research Council
and the Swedish Foundation for Strategic Research (SSF) through the
Gothenburg Mathematical Modelling Centre (GMMC).

\end{document}